\documentclass[twoside,11pt]{article}

\usepackage{tabularx, booktabs, multirow}
\usepackage{authblk}
\usepackage{adjustbox}
\usepackage[a4paper, total={6in, 8.5in}]{geometry}
\usepackage{xcolor, graphicx, pgfplots, pgfplotstable}
\usepackage{amsmath, amsfonts}
\usepackage{natbib}
\usepackage{csquotes}
\usepackage{siunitx, complexity}
\usepackage{bm} 

\DeclareSIUnit{\Bit}{Bit}
\DeclareSIUnit{\Bits}{Bits}

\renewenvironment{abstract}
 {\small
  \begin{center}
  \bfseries \abstractname\vspace{-.5em}\vspace{0pt}
  \end{center}
  \list{}{
    \setlength{\leftmargin}{.4cm}%
    \setlength{\rightmargin}{\leftmargin}%
  }%
  \item\relax}
 {\endlist}

\title{\vspace{-2cm}
    A Rigorous Information-Theoretic Definition of Redundancy and Relevancy in
    Feature Selection Based on (Partial) Information Decomposition
}

\author[1,*]{Patricia Wollstadt}
\author[1]{Sebastian Schmitt}
\author[2]{Michael Wibral}

\affil[1]{Honda Research Institute Europe GmbH, Offenbach/Main, Germany}
\affil[1]{Campus Institute for Dynamics of Biological Networks,
Georg-August University G{\"o}ttingen,
G{\"o}ttingen, Germany}
\affil[*]{Corresponding author: patricia.wollstadt@honda-ri.de}
\date{}

\begin{document}

\maketitle

\begin{abstract}
Selecting a minimal feature set that is maximally informative about a target variable is a central task in machine learning and statistics. Information theory provides a powerful framework for formulating feature selection algorithms---yet, a rigorous, information-theoretic definition of feature relevancy, which accounts for feature interactions such as redundant and synergistic contributions, is still missing. We argue that this lack is inherent to classical information theory which does not provide measures to decompose the information a set of variables provides about a target into unique, redundant, and synergistic contributions. Such a decomposition has been introduced only recently by the partial information decomposition (PID) framework. Using PID, we clarify why feature selection is a conceptually difficult problem when approached using information theory and provide a novel definition of feature relevancy and redundancy in PID terms. From this definition, we show that the conditional mutual information (CMI) maximizes relevancy while minimizing redundancy and propose an iterative, CMI-based algorithm for practical feature selection. We demonstrate the power of our CMI-based algorithm in comparison to the unconditional mutual information on benchmark examples and  provide corresponding PID estimates to highlight how PID allows to quantify information contribution of features and their interactions in feature-selection problems.
\end{abstract}


\section{Introduction}
Which of the many regressor variables in today's large data sets can be used to model or predict an outcome variable, and which variables can be neglected? Answering this question in a process termed feature selection has become a central task in machine learning and statistical modeling, in particular with the increasing availability of large-scale, multivariate data sets. The objective of feature selection is to select a minimal subset of the input variables that provides maximum information about a target variable, often with the goal of minimizing the generalization error in a subsequent learning task, such as classification or regression \citep{Guyon2003,Vergara2014}. A further goal is to reduce training time and improve model performance by increasing interpretability and by reducing effects of the curse of dimensionality \citep{Lensen2018,Guyon2003,Li2018}.

From the requirement that selected feature sets should be minimal, yet maximally informative, it is immediately clear that one does not want to include regressor variables into the feature set if they have information that is already carried redundantly by other variables. Instead, one wants to include variables that have information about the target which they carry uniquely. Additionally, one likes to include synergistic variables, that is multiple variables which only jointly provide the information needed to properly model the outcome.  Such interactions between regressors is well known from classical statistics.

Curiously, even today's most advanced feature selection algorithms that are rooted in information theory lack a way of describing these three possible ways---uniquely, redundantly, synergistically---in which regressor variables carry information about the outcome. This blocks the way to translate the simple intuitions above into working algorithms, and also leads to a lack of conceptual clarity. In fact, even in information theory, the decomposition of the information that a set of variables has about another (outcome) variable into contributions carried uniquely by individual variables, redundantly by multiple variables or that is available only when considering variables jointly has been an open problem until very recently, as laid out in~\cite{Williams2010}. However, this problem, termed partial information decomposition (PID), is now well understood and solutions are available, which finally allows formulating the above intuitions about which variables would be useful features, and which should be discarded, as a rigorously defined information-theoretic problem.

In the present work, we use this novel framework of PID to first clarify why feature selection has been a difficult problem in the past---also from a conceptual point of view. We reanalyze existing feature selection frameworks in the light of their PID and show that approaches based on conditional mutual information (CMI) criteria are strictly preferable over approaches based on the unconditional mutual information and propose a practical iterative CMI-based forward feature selection algorithm based on the improved understanding provided by PID\@. We demonstrate the power of this algorithm on well-known benchmark examples, and also provide the corresponding PID measures, thereby  highlighting how PID leads to a better understanding of the role different variables play as features. We also compare the information-theoretic approach to two established models of feature selection based on sparse linear models, i.e., least angle regression (LARS) \citep{Efron2004}, and random forests (RF) \citep{Breiman2001}.

\section{Background}\label{sec:prior_art}

In feature selection, a general approach is variable filtering, which ranks variables by their relevancy with respect to the target, as measured by a pre-defined criterion. Further approaches include wrapper methods or embedded methods, which optimize the feature set based on the performance of a subsequent or embedded learning algorithm (e.g.,~\citealt{Hastie2009}). In general, filtering is computationally cheaper, less prone to overfitting and provides a \enquote{generic} feature selection approach that is independent of specific, subsequently applied inference models (e.g.,~\citealt{Guyon2003,Duch2006,Hastie2009}). A popular choice for filtering criteria are information-theoretic quantities \citep{Shannon1948,MacKay2005}, such as the mutual information (MI) and the conditional mutual information (CMI, e.g.,~\citealt{Battiti1994,Duch2006,Brown2012}), as well as heuristics derived from both measures \citep{Vergara2014,Guyon2003,Chandrashekar2014,Brown2012}. These quantities are popular because they are inherently model free such that variable dependencies of arbitrary order can be captured, while only minimal assumptions about the data are required for estimation.

Even though, information theory is a popular tool for constructing feature selection criteria, existing definitions of feature relevancy and redundancy (e.g.,~\citealt{John1994,Bell2000}, see~\citealt{Vergara2014} for a review), fail to rigorously capture the notions of feature relevancy and redundancy, in particular for sets of interacting features. As a result, practical feature selection approaches struggle with a clear definition of how interactions between variables contribute to relevancy and redundancy in the data and how interactions influence the selection procedure (e.g.,~\citealt{Brown2012}). We argue that this lack results from the strict definition of MI as the information shared between two variables or two sets of variables \citep{Shannon1948,MacKay2005}, such that interactions between two or more variables can not be described in detail. The decomposition of the joint MI between three or more variables has only recently become possible through the theoretical extensions of classical information theory by the PID framework \citep{Williams2010}, which provides definitions and measures that allow for an unambiguous description of how multiple variables contribute information about a target.

In the following, we will first introduce necessary information-theoretic preliminaries (Section~\ref{sec:info_theory}). We will then highlight where classical information theory lacks in methods to describe multivariate information contribution in feature selection and introduce the PID framework \citep{Williams2010} to close this conceptual gap (Section~\ref{sec:pid_intro}). Using PID, we will go on to define feature relevancy and redundancy in information-theoretic terms in Section~\ref{sec:methods}.

\subsection{Information-Theoretic Preliminaries}\label{sec:info_theory}

In classical information theory\footnote{For a detailed introduction see~\cite{Cover2005} or~\cite{MacKay2005}.} as conceived by~\cite{Shannon1948}, the mutual information (MI) quantifies the information that is shared between two random variables, $X$ and $Y$, as the expected information one variable provides about the other,

\begin{equation*}
    I(X;Y) = \sum_{x\in\mathcal{A}_X, y\in\mathcal{A}_Y} p(x,y)
        \log \frac{p(x,y)}{p(x)p(y)},
    \label{eq:mutual_information}
\end{equation*}

\noindent where $p(x)$ denotes the probability of observing outcome $x$ for variable $X$ and is a shorthand for $p(X = x)$, and $\mathcal{A}_X$ is the support of random variable $X$. Furthermore, $p(x,y)$ denotes the joint probability for simultaneously observing the outcomes $x$ for variable $X$ and $y$ for variable $Y$. The MI is symmetric in $X$ and $Y$ and describes the information which $X$ provides on $Y$ and vice versa. It is always non-negative, it is zero only for independent variables, i.e., $ p(x,y)=p(x)p(y)$, non-zero for any dependence between $X$ and $Y$, and is upper bounded by the entropies, $H(X)=-\sum_{x\in\mathcal{A}_X} p(x)\log p(x)$ and $H(Y)$. The MI may be interpreted as the Kullback-Leibler divergence between the variables' joint distribution, $p(x,y)$, and the assumption of statistical independence, $p(x)p(y)$. Note that in its original form, the MI is strictly defined for two variables or two sets of variables, $\mathbf{X}$ and $\mathbf{Y}$.

To measure the influence a third variable, $Z$, has on the relationship between two variables, $X$ and $Y$, we may calculate the conditional mutual information (CMI), which quantifies the information shared between $X$ and $Y$, given the outcome of $Z$ is known,
\begin{equation}
    I(X;Y|Z) = \sum_{x\in\mathcal{A}_X, y\in\mathcal{A}_Y, z\in\mathcal{A}_Z}
        p(x,y,z) \log \frac{p(x,y|z)}{p(x|z) p(y|z)},
    \label{eq:conditional_mutual_information}
\end{equation}
\noindent where again lower-case letters indicate realizations of random variables and $p(x|z)$ is a shorthand for the conditional probability $p(X=x|Z=z)$. Please note that 
$p(x|y)p(y)=p(x,y)$ holds for conditional probabilities $p(x|y)$.

However, as we will further illustrate below the CMI does not provide us with a detailed account of how $X$ provides information about $Y$ (or vice versa) in the the context of $Z$, but rather quantifies the summarized contribution of $X$ with respect to $Y$ in the context of $Z$. In particular, conditioning on $Z$ may have one of three effects on the MI between $X$ and $Y$: first, the MI remains the same, $I(X;Y|Z)=I(X;Y)$, second, the information $X$ provides about $Y$ \textit{decreases} in the context of $Z$, $I(X;Y|Z)<I(X;Y)$, third, the information $X$ provides about $Y$ \textit{increases} in the context of $Z$, $I(X;Y|Z)>I(X;Y)$. The second and third case are commonly interpreted as $X$ and $Z$ providing primarily redundant information about $Y$, and $X$ and $Z$ providing primarily synergistic or complementary information about $Y$. Note however, that these two contributions may occur simultaneously \citep{Williams2010}, such that the CMI does not allow for an exact quantification of the magnitude of both contributions. Furthermore, the first case, $I(X;Y|Z)=I(X;Y)$, may not be indicative of an independence of $Z$, but may also occur whenever redundant and synergistic contributions cancel each other.

Note that especially the occurrence of synergistic information is often neglected in filtering approaches, which often assume independence between features to allow for easier modeling (e.g.,~\citealt{Dash1997,Hall2000}, see also the review by~\citealt{Brown2012}). Yet, synergistic contributions may occur already in simple settings. As an example, consider a system where an input $X$ has a relationship with a target variable $Y$, but this relationship is corrupted by noise, $Z$ \citep{MacKay2005}. The noise is independent of the input and the target, such that $I(X;Z)=I(Y;Z)=0$. However, the information $X$ provides about $Y$ increases if the noise $Z$ is known, i.e., $I(X;Y|Z)>I(X;Y)$. So adding the noise $Z$ as a feature may add to the explanatory power of $X$ with respect to $Y$, where the contribution of $Z$ may be interpreted as \textit{decoding} the information in $X$ about $Y$ (see~\citealt{Griffith2014} for further examples of synergistic and redundant information in Boolean operations).

Previous approaches have tried to handle the conceptual gap in describing the structure of information contribution of two or more variables about a third by combining MI and CMI terms (e.g.,~\citealt{Watanabe1960,Garner1962,Tononi1994,McGill1954,Bell2003}). For example, approaches in the context of feature selection use various combinations of classical information-theoretic quantities to decompose (joint) variable contributions into relevant and redundant ones (reviewed in~\citealt{Brown2012}). However, these approaches are inherently limited in their ability to provide such a decomposition as we will review in the next section.

\subsection{The Partial Information Decomposition Framework}\label{sec:pid_intro}

The PID framework by~\cite{Williams2010} extends classical information theory by proposing a decomposition of the information two or more variables provide about a third. Formally, we consider the joint MI between a set of assumed inputs, $\{X_1, X_2\}$, and an output, $Y$,
\begin{equation*}
    I(Y; X_1, X_2) =
        \sum_{y \in \mathcal{A}_{Y}, x_1 \in \mathcal{A}_{X_1}, x_2 \in \mathcal{A}_{X_2}}
            p(y, x_1, x_2) \log_2 \frac{p(y|x_1, x_2)}{p(y)},
    \label{eq:multi_mi}
\end{equation*}
\noindent which denotes the total amount of information $\{X_1, X_2\}$ contains about $Y$. Note that again each of the inputs may be a multivariate variable, $\mathbf{X}_i$.

Here, the PID framework by~\cite{Williams2010} proposes to decompose the joint MI into four non-negative contributions, termed \textit{atoms},
\begin{equation}
    \begin{aligned}
        I(Y; X_1, X_2) &= I_{unq}(Y;X_1\setminus X_2) + I_{unq}(Y;X_2\setminus X_1)  \\
        & \qquad   + I_{shd}(Y;X_1, X_2) + I_{syn}(Y;X_1, X_2),
    \label{eq:pid_sum}
    \end{aligned}
\end{equation}
where
\begin{enumerate}
    \item $I_{unq}$ denotes \emph{unique information} provided exclusively by
    either $X_1$ or $X_2$ about $Y$;
    \item $I_{shd}$ denotes \emph{shared information} provided redundantly by both $X_1$ and $X_2$ about $Y$;
    \item $I_{syn}$ denotes \emph{synergistic information} provided jointly by $X_1$ and $X_2$ about $Y$.
\end{enumerate}

\begin{figure}[ht]
    \centering
    \includegraphics[scale=0.5]{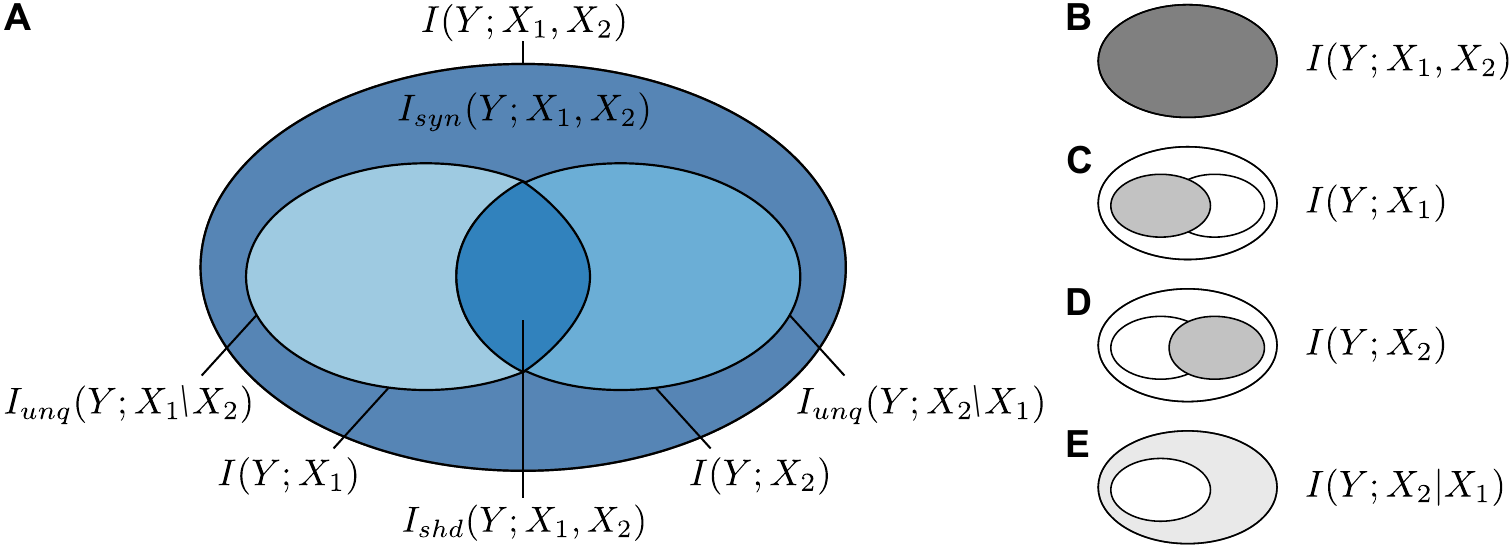}
    \caption{\textbf{Partial information decomposition (PID) framework.}
    \textbf{A} PID diagram of the joint mutual information (MI), $I(Y; X_1, X_2)$, see main text (modified from \protect\citealt{Williams2010}).
    Relation between classical information-theoretic terms and PID\@:
    \textbf{B} Joint MI between $X_1$, $X_2$, and $Y$;
    \textbf{C} MI between $X_1$ and $Y$;
    \textbf{D} MI between $X_2$ and $Y$;
    \textbf{E} conditional mutual information (CMI) between $X_2$ and $Y$ given $X_1$, which corresponds to the unique, $I_{unq}(Y;X_2\setminus X_1)$ and synergistic information, $I_{syn}(Y;X_2, X_1)$.
    }\label{fig:pid_intro}
\end{figure}

\noindent The synergistic information, also termed complementary information, is
information that can only be obtained by considering variables together and can
not be gained from one of the variables alone. The atoms and their relation are shown graphically in Figure~\ref{fig:pid_intro}A.

The PID atoms relate to the MI and CMI between inputs, $X_1$ and $X_2$, and target $Y$ as,
\begin{equation}
    \begin{aligned}
        I(Y; X_1) &= I_{unq}(Y;X_1 \setminus X_2) + I_{shd}(Y;X_1, X_2), \\
        I(Y; X_2) &= I_{unq}(Y;X_2 \setminus X_1) + I_{shd}(Y;X_1, X_2).
        \label{eq:pid_mi}
    \end{aligned}
\end{equation}
and,
\begin{equation}
    \begin{aligned}
        I(Y; X_1|X_2) &= I_{unq}(Y;X_1 \setminus X_2) + I_{syn}(Y;X_1, X_2) \\
        I(Y; X_2|X_1) &= I_{unq}(Y;X_2 \setminus X_1) + I_{syn}(Y;X_1, X_2).
        \label{eq:pid_cmi}
    \end{aligned}
\end{equation}
Note that the five equations~\ref{eq:pid_sum} to~\ref{eq:pid_cmi} are partially redundant and relate three classical information-theoretic terms to four PID atoms. Hence, the system is under-determined and we have to provide an additional definition for at least one of the atoms, such that then all other atoms can be derived via these equations. This also means that neither of the proposed atoms can be derived from measures defined in classic information-theoretic measures, e.g., by subtracting individual terms (Figure~\ref{fig:pid_intro}B--E). Based on this result by~\cite{Williams2010} it can immediately be seen that classical information theory does not allow for the quantification of the individual contributions as would be desirable in applications such as feature selection.

\cite{Williams2010} provide an axiomatic definition of the shared information together with a corresponding measure, $I_{min}$. However, the proposed axioms do not uniquely determine a measure of each of the atoms and subsequent work has proposed further PID measures and axioms \citep{Griffith2014,Bertschinger2014,Harder2013,Makkeh2021,Gutknecht2021}, which differ subtly in their operational interpretation. At the time of writing, one of the most popular approaches to quantify PID is the one proposed by Bertschinger, Rauh, Olbrich, Jost, and Ay (BROJA measure,~\citealt{Bertschinger2014}), who introduced a measure of unique information based on game-theoretic considerations, and which shares its axiomatic foundation with the measures proposed by~\cite{Griffith2014}. See~\cite{Bertschinger2014} and~\cite{Griffith2014} for the PID of some canonical examples and differences in estimates for different PID measures.

We here apply the BROJA measure because its definition appeals to multiple variables \enquote{competing} for an explanation of the output variable. The authors derive a definition of the unique information from the argument that truly unique information should be exploitable in a decision problem. Based on this assumption, the authors argue that the unique information should depend only on the marginal probability distributions, $p(Y,X_1)$ and $p(Y,X_2)$, but not the full joint distribution, $p(Y,X_1,X_2)$. Thus, the unique information should not change for different joint distributions, $q(Y,X_1,X_2)$, from a space $\Delta_P$ with the same marginals as $p(Y,X_1,X_2)$,
\begin{equation}
    \begin{split}
    \Delta_P = \{ q \in \Delta : q(Y=y,X_1=x_1)=p(Y=y,X_1=x_1) \\
                        \text{ and } q(Y=y,X_2=x_2)=p(Y=y,X_2=x_2) \},
    \end{split}
\end{equation}
where $\Delta$ is the space of all probability distributions over the support of $Y$, $X_1$, and $X_2$. From their assumption, Bertschinger et al.\ derive the unique information as
\begin{equation}
    I_{unq}(Y;X_1\setminus X_2) = \min_{q \in \Delta_P} I_q(Y;X_1|X_2),
\end{equation}
where $I_q(Y;X_1|X_2)$ is a conditional mutual information computed with respect to $q$. The measure can be estimated from data by means of convex optimization \cite{Makkeh2018}. Note that the BROJA PID measure is defined for the case of two input variables only. However, this is sufficient for an application to iterative forward feature selection as illustrated in the next section. Furthermore, the estimator used here is applicable to discrete data only, which we deemed sufficient for the applications shown here. Estimators for continuous data have been developed in more recent work~\cite{Schick-Poland2021}. Note that both restrictions concern the estimation of PID from data, while the definition of feature relevance proposed in the next section hinges on the definitions of PID atoms as introduced by~\cite{Williams2010} only.

\section{Methods}\label{sec:methods}

Based on the introduced PID framework, we will present a rigorous definition of variable relevancy and redundancy in settings for which feature-independence can not be generally assumed (Section~\ref{sec:pid_def}). Based on this definition, we will show that using the CMI as a filtering criterion in forward feature selection algorithms maximizes feature relevancy, while simultaneously minimizing redundant contributions, and we provide a theoretical justification of why the CMI generally outperforms the MI as a selection criterion (Section~\ref{sec:theoretical_guarantees}). To overcome the practical problems commonly encountered when estimating information-theoretic quantities from data, we propose the use of a recently introduced sequential forward-selection algorithm (Section~\ref{sec:cmi_algo}). In Section~\ref{sec:experiments}, we will present results from applying both the proposed algorithm and PID estimation to benchmark models from literature, where we demonstrate that the CMI criterion generally outperforms the MI criterion, and that PID allows for a quantitative description of feature interactions in terms of synergistic and redundant contributions.

\subsection{Using PID to Define Feature Relevancy and Redundancy}\label{sec:pid_def}

We assume a feature selection problem with a target variable, $Y$, and a set of $N$ input variables, $\mathbf{X}=\{X_k\}_{k=1}^N$, from which we want to select a set of relevant features, $\mathbf{S} \subseteq \mathbf{X}$. We wish to define feature relevancy such that, intuitively, a variable is considered relevant if it i) uniquely provides information about the target, ii) provides information in the context of other variables, and iii) does not provide information that is redundant with information already contained in the feature set. As a result, a filtering criterion for including a feature should return a high score if the first two conditions are met, and should be low if the last is not met. Note that in practice, finding an optimal feature set, $\mathbf{S}$, i.e., the set of features that provides maximum information about $Y$ while having minimal cardinality, is computationally not feasible already for small input sizes $N$. It requires the evaluation of all possible $2^N$ subsets, $\mathbf{S}' \subseteq \mathbf{X}$, i.e., the power set of the $N$ input variables. Indeed, selecting the optimal feature set is an \NP-hard problem \citep{Amaldi1998}. Hence, we provide a definition applicable to an approximative feature selection procedures such as sequential forward-selection or backward-elimination \citep{Hastie2009}.

PID allows for an immediate quantification of the three properties stated above in information-theoretic terms:
At a feature selection iteration $i$ we assume there exists the set of already identified relevant features, $\mathbf{S}_i=\{F_j\}_{j=0}^{i-1}\subseteq \mathbf{X}$. The relevancy of an additional feature, $F \in \mathbf{X}\setminus\mathbf{S}_i$, is then given by the sum of the information provided uniquely by the feature, $I_{unq}(Y;F \setminus \mathbf{S}_i)$, and the information provided synergistically by the feature and the already selected feature set, $I_{syn}(Y;F, \mathbf{S}_i)$. Finally, the redundancy of $F$ with respect to the already selected feature set is quantified by the shared information, $I_{shd}(Y;F, \mathbf{S}_i)$.

\begin{figure}[ht]
    \centering
    \includegraphics[scale=0.5]{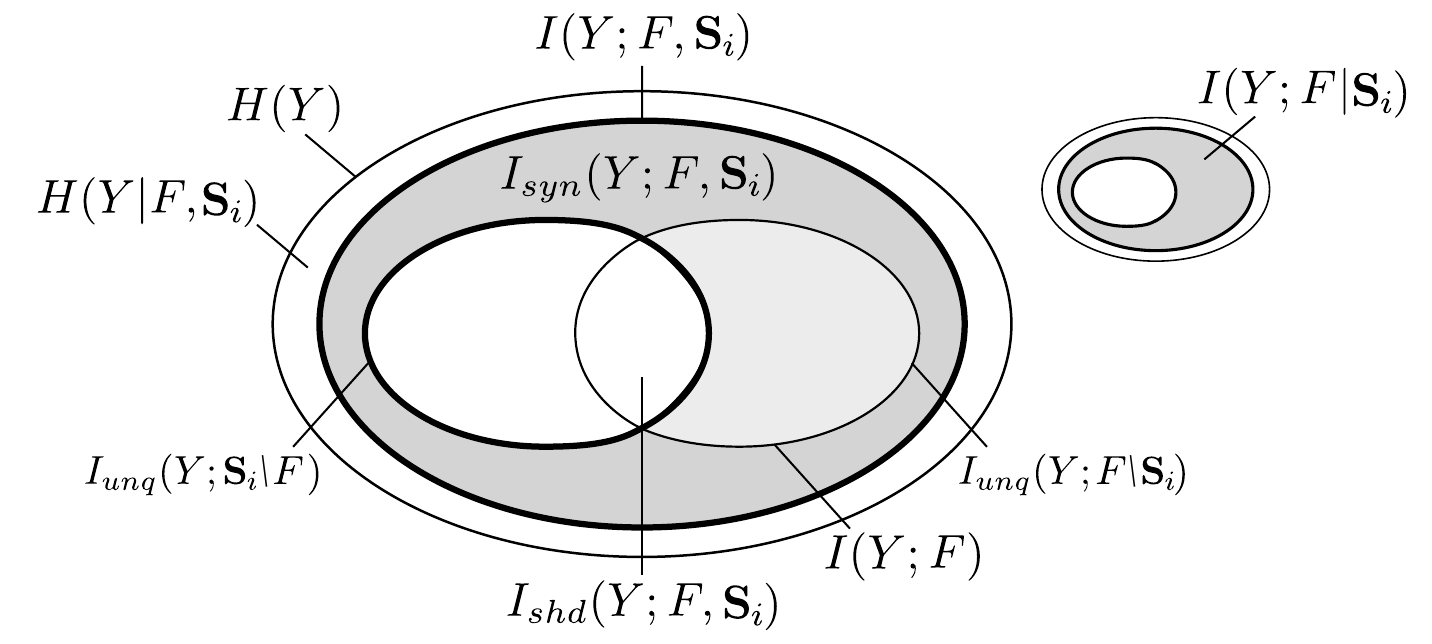}
    \caption{\textbf{Partial information decomposition (PID) diagram of feature relevancy.}
    PID diagram of the conditional mutual information (CMI) criterion for feature selection, $I(Y;F|\mathbf{S}_i)$, which may be decomposed into the unique information $I_{unq}(Y;F\setminus \mathbf{S}_i)$ and the synergistic information $I_{syn}(Y;F,\mathbf{S}_i)$, while not including the redundancy between $F$ and $\mathbf{S}_i$. See main text.
    }\label{fig:pid_cmi}
\end{figure}

Comparing this definition of relevancy to Equation~\eqref{eq:pid_cmi}, we can directly see that it is equivalent to the CMI\@. Hence, we define the criterion whether to include a feature into the relevant set of selected features as
\begin{equation}
    I(F; Y|\mathbf{S}_i) = I_{unq}(Y;F \setminus \mathbf{S}_i) + I_{syn}(Y;F, \mathbf{S}_i),
    \label{eq:pid_relevance}
\end{equation}
where $i$ indicates either the $i$-th iteration in a step-wise algorithm or the rank of the feature, and $\mathbf{S}_i$ is the set of selected features at the $i$-th iteration or of rank $i$, respectively (see Figure~\ref{fig:pid_cmi} for the corresponding PID diagram). From our PID-based definition, we see that the CMI criterion returns a high score for features that either have a high unique contribution, or a high synergistic contribution with respect to the currently selected feature set, $\mathbf{S}_i$, or provide both contributions. On the other hand, the score is low if a variable carries primarily information that is already redundantly present in $\mathbf{S}_i$ and which does not contribute to the CMI\@. Hence, the CMI is able to capture relevancy resulting from variable interactions, where it only considers the \textit{partial} contribution of the feature about the target that is relevant.

Our definition of feature relevancy does not hinge on a concrete measure of PID but rather on the definition of PID atoms as provided in the original publication by~\cite{Williams2010}. Hence, while the choice of the concrete PID measure and estimator influences the values estimated from data, results should be qualitatively similar for all measures adhering to the axioms defined by Williams and Beer.

For completeness, we also express the MI in terms of our definition of relevancy and redundancy,
\begin{equation}
    I(Y; F) = I_{unq}(Y;F \setminus \mathbf{S}_i) + I_{shd}(Y;F, \mathbf{S}_i).
    \label{eq:pid_mi_criterion}
\end{equation}
\noindent We see that the MI criterion---opposed to the CMI---misses the synergistic information contribution between $F$ and $\mathbf{S}_i$. Furthermore, the MI also includes the redundant information, $I_{shd}(Y;F, \mathbf{S}_i)$. Hence, using the PID framework, we can immediately show that the CMI is able to account for synergistic interactions between variables, while not considering redundant information. In contrast, the MI criterion only considers unique contributions and fails to include synergistic contributions and potentially fails to remove redundancies in the feature set. We will give a detailed account of the advantage of the CMI as a selection criterion over the MI in the next section, including a simple example in Section~\ref{sec:algo_toy_example}, and we will demonstrate the consequences of choosing either criterion in feature selection in the experiments.

\subsection{Theoretical Guarantees and Extension to PID for more than Three Variables}\label{sec:theoretical_guarantees}

As introduced above, for two features, $F_1$ and $F_2$ which provide information about a target, $Y$, either the MI, $I(Y;F_2)$, or the CMI, $I(Y;F_2|F_1)$ can be larger. Therefore, it is not intuitively obvious which is the better choice as a selection criterion in the context of iterative feature selection. However, based on theoretical arguments  provided in this work using PID we show that it is indeed generally better to use CMI in iterative feature selection, given it can be robustly estimated from the available data. The reason for this is that the information which leads to the MI of a feature $F_i$, to be larger than the CMI of that feature (conditioned on the set of already included features) necessarily needs to be redundant with the already included feature set. Conversely, the information that leads to the CMI being larger than the MI, is necessarily synergistic in nature and cannot be accessed by the MI alone.

To see this, we start considering the case of two features $F_1$, $F_2$, and assume without loss of generality that $I(Y;F_1)>I(Y;F_2)$. Thus, when considering the inclusion of the first variable by MI or CMI, where we condition on the information carried by the empty set of features, $F_1$ will be included first. Comparing now the two metrics for the inclusion of the next variable we find that:
\begin{equation}
I(Y;F_2)-I(Y;F_2|F_1)=I_{shd}(Y;F_2,F_1)-I_{syn}(Y;F_2,F_1).
\end{equation}
\noindent From this, we immediately see that $I(Y;F_2)$ can only be larger than $I(Y;F_2|F_1)$ by virtue of a large $I_{shd}(Y;F_2,F_1)$, i.e., by information already provided by the first variable $F_1$; in such a case one would be ill-advised to include $F_2$ into the feature set, despite its large MI\@. In contrast, the term that would make $I(Y;F_2|F_1)$ larger than $I(Y;F_2)$ is the synergy $I_{syn}(Y;F_2,F_1)$\@. This synergy indeed represents information not accessible by $F_1$ alone. A larger CMI, and thereby a large synergy, is thus indeed a reason to include $F_2$\@. In sum, a decision on whether to include $F_2$ should be based on $I(Y;F_2|F_1)$, not on $I(Y;F_2)$.

The above argument naturally extends to larger feature sets and iterative inclusion by replacing $F_1$ and $F_2$ by the appropriate features under consideration. In this sense, we need not necessarily consider a PID for more than three variables (two input and one output variable). For a deeper understanding of the feature selection problem one may consider the full structure of multivariate MI in terms of conditional and pairwise MI terms, and the underlying many-variable PID\@. In Appendix~A, we introduce the structure of the PID problem for the multivariate case.

\subsection{An Algorithm for Sequential Forward-Selection Using a CMI-Criterion}\label{sec:cmi_algo}

To perform CMI-based feature selection in a computationally feasible fashion, we propose to use a sequential forward-selection algorithm that was recently introduced in the context of network inference from multivariate time series data \citep{Novelli2019,Wollstadt2019}. We here consider a forward-selection approach to avoid estimating the CMI in too high-dimensional spaces, which would be required for approaches such as pure backward-elimination or testing each variable by conditioning on the whole remaining variable set. We briefly introduce the algorithm, its implementation, and the estimation of the CMI from data in this Section. For a more detailed account on the technical aspects of the algorithm, including proofs and an empirical evaluation refer to~\cite{Novelli2019}.

\subsubsection{Sequential Forward-Selection Algorithm}

The algorithm starts from an empty feature set, $\mathbf{S}_0=\emptyset$, and the full set of variables, $\mathbf{X}_0=\mathbf{X}$, and includes in each step a feature using the CMI criterion of Equation\ \ref{eq:conditional_mutual_information},
\begin{equation}
    F_i = \max_{X \in \mathbf{X}_i} I(X; Y|\mathbf{S}_i),
    \label{eq:cmi_criterion}
\end{equation}
where $\mathbf{X}_i$ denotes the remaining input variables in iteration $i$, and $\mathbf{S}_i$ the set of already selected features. The feature with the largest contribution, $F_i$, is then included into the feature set and removed from the set of remaining variables,
\begin{equation}
    \begin{aligned}
        \mathbf{S}_{i+1} &= \mathbf{S}_i \cup F_i \, , \\
        \mathbf{X}_{i+1} &= \mathbf{X}_i \setminus F_i\, .
    \end{aligned}
    \label{eq:inclusion_update}
\end{equation}
Hence, each feature is evaluated in the context of \textit{all} already selected features such that also higher-order interactions are accounted for.

When using the CMI or MI as a criterion in feature selection, it is central to determine whether the criterion is truly non-zero. While in theory the criterion is zero for (conditionally) independent variables, in practice estimators may return non-zero results also for independent variables due to finite sample size \citep{Paninski2003,Kraskov2004,Hlavackova-Schindler2007}. The proposed algorithm handles this practical estimation problem by performing non-parametric statistical tests against surrogate data to assess whether the estimate is statistically significant under the Null hypothesis of conditionally independent variables \citep{Novelli2019}. To generate the Null distribution, the CMI is repeatedly estimated from surrogate data, which is generated by permuting the data such that the joint distribution is destroyed while the marginal distributions stay intact. We here use a testing procedure introduced and described by~\cite{Novelli2019}, which has been shown to effectively control the family-wise error rate over multiple tests and prevent an increased false-positive rate \citep{Novelli2019,Wollstadt2019}. A short description of the testing procedures is provided in Appendix~B, while a thorough empirical evaluation is presented in~\cite{Novelli2019}, as well as a proof of the equivalence of the testing procedure and classical statistical correction methods.

The inclusion terminates once $\max_{X \in \mathbf{X}_i} I(X;Y|\mathbf{S}_i)$ is no longer statistically significant, i.e., if none of the remaining variables adds information about the target in a significant fashion. The algorithm then performs a backward-elimination where iteratively the weakest feature, given the full selected feature set, is tested for a statistically significant contribution to the target, $Y$ \citep{Novelli2019}. The weakest feature here denotes the feature with the smallest MI about $Y$ given the remaining feature set, $\min_{F \in \mathbf{S}} I(F;Y|\mathbf{S}\setminus F)$. The feature is excluded if it no longer provides significant information in the context of the full feature set. If the contribution of the weakest feature is significant, the backward-elimination terminates and the final feature set is returned. This \enquote{pruning} of the feature set ensures that the identified set is minimal while providing a maximum of statistically significant information about the target \citep{Lizier2012}. Removing redundant sources in an \emph{iterative} fashion thereby prevents the removal of all sources providing the same, redundant information. See also \cite{Runge2015,Novelli2019,Sun2014,Sun2015} for discussions of the pruning step for iterative feature selection.

Note that the algorithm may also be used with a MI criterion for selecting the features,
\begin{equation}
    F_i = \max_{X \in \mathbf{X}_i} I(X; Y),
    \label{eq:mi_criterion}
\end{equation}
\noindent where all other steps remain identical.

\subsubsection{Implementation and Estimation from Data}\label{sec:algo_implementation}

We use an implementation of the proposed approach provided as part of the IDTxl Python toolbox \citep{Wollstadt2019} which makes use of estimators implemented in the JIDT toolbox \citep{Lizier2014jidt}. IDTxl provides estimators for various use cases such as discrete and continuous data, as well as model-free estimators and estimators for jointly Gaussian variables. To estimate the CMI and MI, we here use a model-free plug-in estimator for discrete data \citep{Hlavackova-Schindler2007} and a model-free nearest-neighbor-based estimator for continuous data by Kraskov et al.~\citep{Kraskov2004}. We chose a discrete estimator to make estimated MI terms comparable to PID terms estimated with the BROJA estimator for discrete data. Plug-in estimators calculate information-theoretic quantities from empirical distributions obtained from data, while the Kraskov estimator uses a nearest-neighbor-based approach, which has been shown to have more favorable bias properties and to be applicable also in higher-dimensional spaces \citep{Kraskov2004,Khan2007,Lizier2014jidt,Xiong2017}. The estimator is therefore particularly suited for application in sequential feature selection \citep{Francois2007,Doquire2012}.

\subsubsection{Example System Illustrating the Algorithm and Application of PID Estimation in Feature Selection}\label{sec:algo_toy_example}

To illustrate the algorithm on a simple toy system, we assume the following system:
%
\begin{equation*}
    \begin{aligned}
        Y &= \sin(\xi_1) + 0.1 \eta_Y \\
        X_1 &= \xi_1 + 0.1 \eta \\
        X_2 &= 0.8 \xi_1 + (1-0.8) \xi_2 + 0.01 \eta,
    \end{aligned}
\end{equation*}
where $\xi_1$, $\xi_2$, $\eta$, $\eta_Y$ are drawn randomly and independently from a standard normal distribution $\mathcal{N}(\mu=0,\sigma=1)$. The example is designed such that both variables $X_1$ and $X_2$ are informative about $Y$, while all information $X_2$ provides is also redundantly present in $X_1$, and $X_2$ provides less information about $Y$ than $X_1$.  Furthermore, the noise source, $\eta$, provides synergistic information in combination with either variable, while by itself being independent of $Y$.

\begin{figure}[ht]
    \centering
    \includegraphics[scale=1.0]{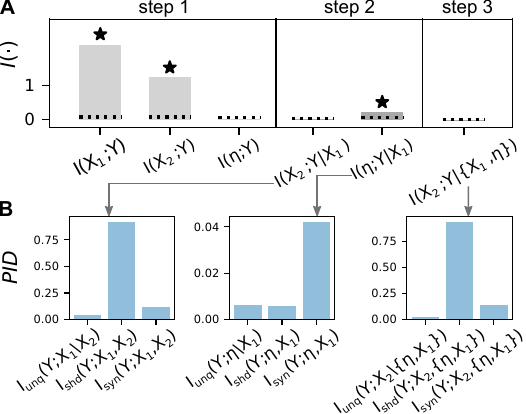}
    \caption{\textbf{Sequential forward feature selection for 4-way PID example system.}
    \textbf{A} Mutual information (MI) and conditional mutual information (CMI) estimates in each selection step, stars indicate statistically significant MI and CMI estimates (see main text).
    \textbf{B} Pairwise partial information decomposition (PID) terms in inclusion steps 2 and 3 for the respective remaining input variables.
    }\label{fig:3way_pid_example}
\end{figure}

Figure~\ref{fig:3way_pid_example} illustrates feature selection for input variables, $\mathbf{X}=\{X_1, X_2, \eta\}$ and target, $Y$, using the proposed algorithm with the CMI selection criterion. In step 1, the MI between each input variable and $Y$ is calculated and tested for statistical significance using the maximum statistic. Here, only contributions $I(X_1;Y)$ and $I(X_2;Y)$ are significant with $I(X_1;Y)>I(X_2;Y)$, leading to an  inclusion of $X_1$ into the feature set according to Equation~\ref{eq:inclusion_update}. In step 2, the CMI between the two remaining variables, $X_2$, $\eta$, conditional on the feature set, $\mathbf{S}_{2}=\{X_1\}$, is calculated, where only $\eta$ shows a significant contribution and is included. In Step 3, only $X_2$ remains to be evaluated and does not provide any significant contribution given the feature set, $\mathbf{S}_{2}=\{X_1,\eta\}$, which leads to a termination of the forward-selection. In the subsequent backward elimination, the weakest feature still contributes information in a statistically significant fashion, which leads to the final feature set $\mathbf{S}=\{X_1,\eta\}$.

Figure~\ref{fig:3way_pid_example}B shows the PID between each currently considered variable, the current feature set, and the target for inclusion steps 2 and 3. In step 2, variable $X_2$ provides information that is mainly redundant with the contribution of variable $X_1$, which is already in the feature set, while $\eta$ provides almost exclusively synergistic information in the context of $X_1$. The CMI correctly identifies $\eta$, thus accounting for synergistic contributions, while not including $X_2$, due to the redundancy in $X_1$ and $X_2$. Note that in step 1, $\eta$ did not provide any information about $Y$ in isolation ($I(\eta;Y)$). In the final step 3, the remaining variable $X_2$ provides information that is redundant with the information provided by the already selected features, resulting in a low and non-significant CMI\@.

In Appendix~C, we show results from applying the algorithm to feature selection problems of larger  sizes relevant for real-world applications.

\section{Experiments and Results}\label{sec:experiments}

We demonstrate our proposed algorithm in a series of experiments on synthetic data obtained from benchmark models from literature. We chose models that were explicitly designed to introduce both synergies and redundancies into the data and used both a CMI (Equation~\ref{eq:cmi_criterion}) and MI (Equation~\ref{eq:mi_criterion}) criterion to illustrate how variable interactions affect feature selection. Additionally, we estimate PID for each experiment to quantify multivariate contributions to feature relevancy and redundancy.

For comparison, we also estimated the importance of the features with alternative established methods. We tested several linear regression models such as least absolute shrinkage and selection operator (Lasso, see, for example~\citealt{Tibshirani1996}), least angle regression (LARS, see, for example \citealt{Efron2004}), and linear support vector regression (see, for example,~\citealt{Smola2004} or~\citealt{bishopPattern2006}). These models include a $L_1$ regularization term on the regression coefficients in order to produce sparse models. The absolute magnitude of the regression coefficients can then be used as a measure of the importance of the corresponding feature. Since our tests showed no systematic differences between these linear approaches on our benchmark data sets, we here only report results for LARS\@. Additionally, we used random forest (RF) regression for feature selection (see, for example,~\citealt{Breiman2001}) as a standard non-linear approach. The feature importance in a RF regression model was estimated by the decrease in model accuracy when single feature values were randomly shuffled \citep{Breiman2001}. The importance values estimated by such methods provide a ranking of the features and allow for selecting the most important features by setting some thresholds on the (relative) importances.

Finally, we also used the above models in sequential forward-selection (see, for example,~\citealt{FERRI1994403}). Here, in an initial step one regression model is trained for each variable as the only input and the variable of the best performing model is selected. In each subsequent iteration step, the feature set is extended by one variable at a time, where the best performing regression model determines the next variable to be added from all possible extensions with one additional variable. A major drawback of the plain forward-selection method is that the number of features to select needs to be provided as an input to the iterations, whereas in practice the number of relevant features is typically unknown. Note, however, that a statistical test of the improvement in accuracy with including a feature is also possible, similar to the sequential forward-selection algorithm proposed here.

The features selected by the forward-selection approaches do not necessarily correspond to the importance values calculated from the respective regression models. The reason is due to the fact that sequential forward-selection is an iterative approach that includes one feature at a time whereas the importance values are obtained from the full regression model and thus always consider all features simultaneously.

We used the scikit-learn Python toolbox \citep{scikit-learn} implementation of all methods for all our experimental results. To make results comparable across methods, we report normalized feature importances. Each experiment was repeated \num{100} times if not specified otherwise and used $N=1000$ samples for estimation. For significance testing, we used a critical alpha level, $\alpha=0.05$. To estimate PID using the BROJA measure, we use the estimator proposed by~\cite{Makkeh2018} as implemented in the IDTxl toolbox \citep{Wollstadt2019}. For continuous features, we discretized the data into $n=5$ bins if not specified otherwise.

\subsection{Experiment I\@: Nested Spheres}\label{sec:experiment_spheres}

\begin{figure}[ht]
    \centering
    \includegraphics{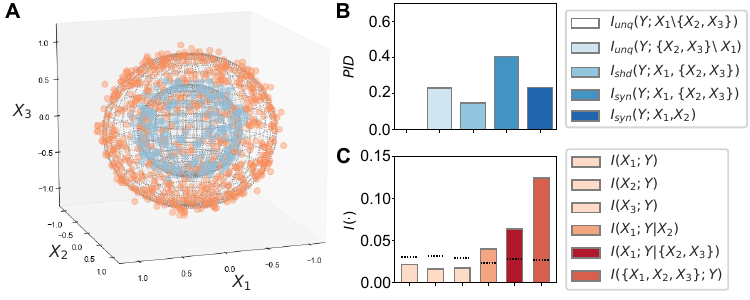}
    \caption{\textbf{Results Experiment I.} Mutual information (MI), conditional mutual information (CMI), and partial information decomposition (PID) terms for three variables sampled from two nested spheres.
    \textbf{A} Data generation, each sphere indicates one class (orange and blue markers);
    \textbf{B} PID terms;
    \textbf{C} MI and CMI terms (dashed horizontal lines indicate significance thresholds).
    }\label{fig:spheres_pid}
\end{figure}

To illustrate a simple case of synergistic information contribution and its quantification using PID, we sampled data from two nested spheres, where each sphere denoted one class, $Y \in \{0,1\}$, and each sample's Cartesian coordinates were considered the input variables, $\mathbf{X} =\{X_1, X_2, X_3\}$  (Figure~\ref{fig:spheres_pid}A). To be able to estimate MI terms using continuous estimators for MI and CMI \citep{Kraskov2004}, we added random Gaussian noise to class labels.

As expected, we observed predominantly synergistic contributions between the inputs, while there was no unique information in individual input variables (Figure~\ref{fig:spheres_pid}B). When considering sets of two variables there was an increase in unique information, accompanied by shared information and a higher synergistic contribution compared to the synergistic contribution of two inputs alone. Accordingly, the pairwise MI, $I(X_i;Y)$, did not show a significant information contribution of individual variables (Figure~\ref{fig:spheres_pid}C), while we found significant CMI, $I(X_i;Y|X_j), X_i, X_j \in \mathbf{X}$, for individual input variables, conditional on at least one other variable. When conditioning on a single variable the CMI was higher than the pairwise MI and was highest when conditioning on both remaining variables.

\subsection{Experiment II\@: Statistical Models}

As a more complex example of both shared and synergistic information contribution, we generated data from two simple statistical models that each combined two input variables, $X_1$ and $X_2$, either via addition or multiplication, while the influence of $X_2$ was varied by a weighting parameter, $\alpha$,
\begin{equation}
    Y = \alpha X_1 + (1-\alpha) X_2 + \sigma \mathcal{N}(0,1), \quad\text{(additive model)}
    \label{eq:stats_model_add}
\end{equation}
\noindent and
\begin{equation}
    Y = X_1^\alpha X_2^{1-\alpha} + \sigma \mathcal{N}(0,1), \quad\text{(multiplicative model)}
    \label{eq:stats_model_mult}
\end{equation}
where $\sigma \mathcal{N}(0,1)$ denotes the addition of Gaussian noise with $\sigma=\in \{0.1, 1.0, 10.0\}$ and $\alpha\in [0, 2]$. Input data were sampled from different distributions as described in Table~\ref{tab:stats_models_params}.

\begin{table}[ht]
    \centering
    \begin{tabular}{lcc}
      \toprule
      Distribution & Formula & Parameter values \\ \midrule
      Uniform     & $f(k;b,a) = \frac{1}{b-a}$ & $b=1$, $a=2$ \\
      Binomial    & $f(k;n,p) = \binom{n}{k} p^k q^{n-k}$ & $n=20$, $p=0.5$, $q = 1-p$ \\
      Poisson     & $f(k;\lambda) = \frac{\lambda^k e^{-\lambda}}{k!}$ & $\lambda_1=4$, $\lambda_1=10$ \\
      Exponential & $f(k;\lambda) =\lambda^k e^{-\lambda}$ & $\lambda = 1.5$ \\ \bottomrule
    \end{tabular}
    \caption{Distributions and parameters used for generating input data in Experiment II.}\label{tab:stats_models_params}
\end{table}

\begin{figure}[ht]
    \centering
    \includegraphics{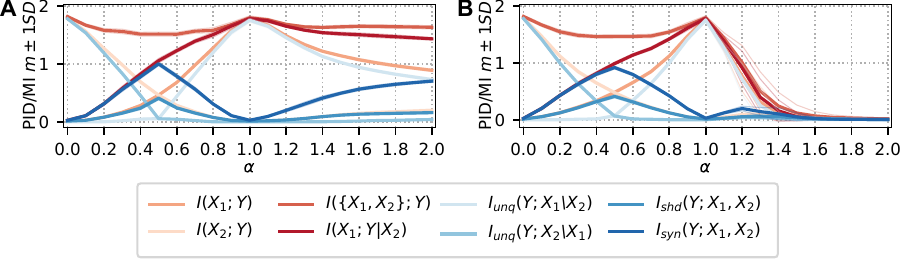}
    \caption{\textbf{Results Experiment II.} Mutual information (MI), conditional mutual information (CMI), and partial information decomposition (PID) terms for (\textbf{A}) additive and (\textbf{B}) multiplicative models with noise level $\sigma=0.1$. Input data were sampled from a uniform distribution (see main text).
    }\label{fig:stats_models_pid}
\end{figure}

Estimated quantities changed as a function of signal-to-noise ratio. While increasing the noise level reduced the absolute CMI, MI, and PID values, results remained qualitatively similar results for different input distributions and noise levels (not shown).
Figure~\ref{fig:stats_models_pid} shows exemplary results for the uniform distribution and noise level $\sigma=0.1$.

For all models, synergy and redundancy reached a local maximum for $\alpha=0.5$, i.e., an equal contribution of both features, and a local minimum for $\alpha=1$, i.e., no contribution of $X_2$. Complementary to synergistic and shared information, the unique information provided by $X_2$ was highest for $\alpha=0$ and was highest for $X_1$ for $\alpha=1$, i.e., whenever the outcome $Y$ was predominantly determined by a single feature only. For the additive models, all contributions increased for larger $\alpha$, while for multiplicative models, contributions decreased.

In sum, both additive and multiplicative combination of input variables introduced synergistic and shared information contributions. This is in line with results by~\cite{Griffith2014} who found that logical \texttt{XOR} leads to higher synergy than logical \texttt{AND}, which corresponds to addition and multiplication of Boolean variables, respectively. The marginal distributions of individual variables had no effect on the interaction of both variables. The experiment illustrates the introduction of synergy and redundancy already through simple algebraic operations, where for $\alpha=1$ synergy and redundancy was of similar magnitude in both operations.

\subsection{Experiment III\@: Friedman Models}\label{sec:experiments_friedman}

\cite{Friedman1991} proposed three regression models that are frequently used to evaluate feature selection algorithms on regression problems, due to their synergistic interactions between input variables (e.g.,~\citealt{Tsanas2010} or a simplified version of model I in~\citealt{Francois2007}). Data were generated using the implementation in the scikit-learn Python toolbox \citep{scikit-learn}.

Friedman model I generates the output variable, $Y$, according to
\begin{equation*}
    Y = 10 \sin(\pi X_1 X_2) + 20 (X_3 - 0.5)^2 + 10 X_4 + 5 X_5 + \sigma \mathcal{N}(0, 1),
    \label{eq:friedman_1}
\end{equation*}
where the inputs  $X_i$ are independently sampled from a uniform distribution over 
$[0,1]$.

For Friedman models II and III we use,
\begin{align*}
    Y &= \sqrt{X_1^2 + (X_2 X_3  - 1 / X_2 X_4)^2}
        + \eta_2 \textrm{, with } \eta_2 \sim \mathcal{N}(0,1), \\
    Y &= \arctan\left(\frac{X_2 X_3 - 1 / (X_2 X_4)}{X_1}\right)
        + \eta_3 \text{, with } \eta_3 \sim \mathcal{N}(0,1),
    \label{eq:friedman_2_3}
\end{align*}
respectively, where the inputs, $X_i$, are uniformly sampled from the intervals
\begin{align*}
0 \leq & X_1 \leq 100, \\
40 \,\pi \leq& X_2 \leq 560\,\pi, \\
0 \leq& X_3 \leq 1, \\
1 \leq& X_4 \leq 11.
\end{align*}
For model I (model II and III), we added five (six) uniformly-distributed nuisance variables, $Z_i$, that were independent of $Y$, to test for the algorithm's ability to discard irrelevant features. Again, we applied the forward-selection algorithm using a CMI and MI criterion, respectively, and estimated selected MI, CMI, and PID terms as in previous experiments.

\begin{figure}[p]
    \centering
     \includegraphics[width=0.97\textwidth]{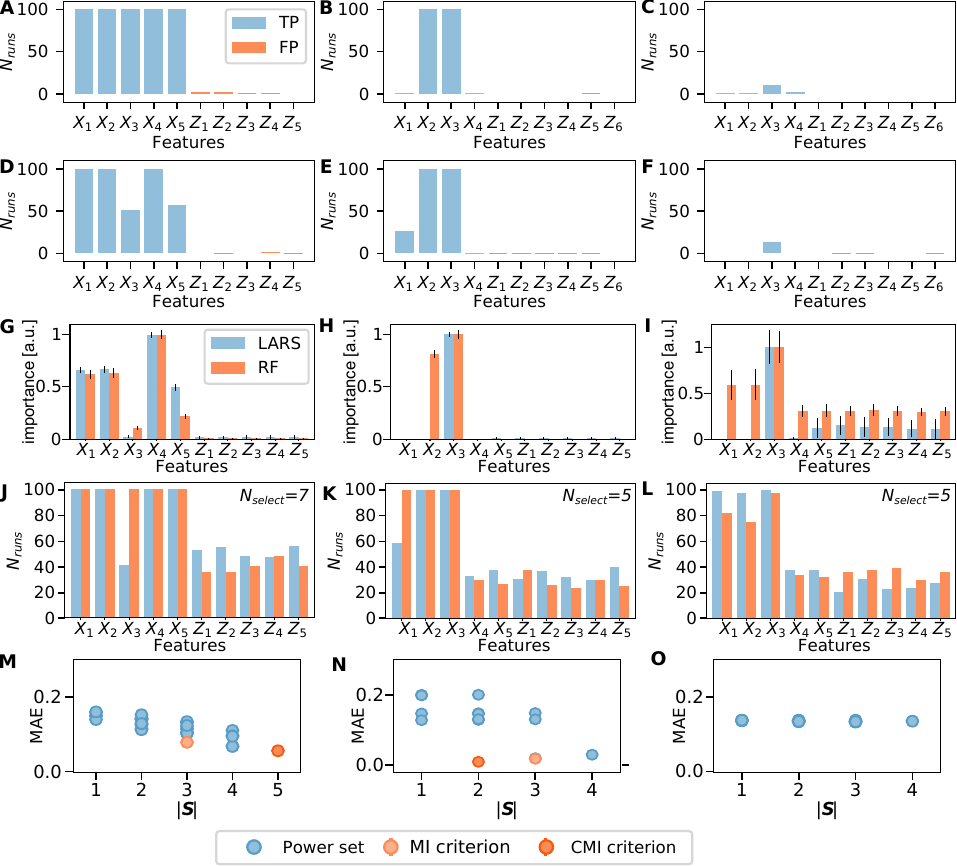}
    \caption{\textbf{Results Friedman models.}
     Variables selected by the proposed forward-selection algorithm using the conditional mutual information (CMI)  for model I (\textbf{A}), model II (\textbf{B}), and model III (\textbf{C}), and the  mutual information (MI) for model I (\textbf{D}), model II (\textbf{E}), and model III (\textbf{F}) as inclusion criterion (from 100 simulation runs). Blue bars indicate correctly included true features (TP\@: true positives), orange bars indicate erroneously included spurious features (FP\@: false positives).
    Feature importance values estimated with LARS and RF regression models for  model I (\textbf{G}),  model II (\textbf{H}) and  model III (\textbf{I}).
    Selected features using sequential forward-selection with LARS and RF regression models for  model I with $N_{select}=7$ selected features (\textbf{J}),  model II with $N_{select}=5$ selected features (\textbf{K}) and  model III with $N_{select}=5$ selected features (\textbf{L}).
    Mean absolute error (MAE) in $k=5$-nearest neighbor prediction of $Y$ from all possible variable subsets of sizes $|\mathbf{S}|$, light and dark orange markers indicate feature sets selected using the MI and CMI criterion respectively. MAE for model I (\textbf{M}), model II (\textbf{N}), and model III (\textbf{O}, here, no feature set was reliably identified by either criterion).
    }\label{fig:friedman_rates_knn}
\end{figure}

For model I, relevant features were reliably detected by the CMI criterion, while false positive rates were well below the critical $\alpha$-level (Figure~\ref{fig:friedman_rates_knn}A, Table~\ref{tab:friedman_rates}). Using the MI criterion, true positive rates dropped mostly because variables $X_3$ and $X_5$ were not selected in many instances (Figure~\ref{fig:friedman_rates_knn}D), while false positive rates increased and where higher than the critical $\alpha$-level.
All tested traditional feature selection methods gave qualitatively similar results. Figure~\ref{fig:friedman_rates_knn}G shows feature importance values for the LARS and RF method. Both methods correctly assigned very low importance to the nuisance variables, $Z_i$, and showed higher importance for $X_i$, while $X_4$ always had highest importance, followed by $X_1$ and $X_2$, and then $X_5$ and $X_3$. While for RF the importance of $X_3$ was small but non-negligible, LARS failed to assign a noticeable importance, but produced values on the same level as the nuisance variables $Z_i$. Thus, importance values assigned by both models resembled results obtained by using the MI selection criterion, with low scores for $X_3$ and $X_5$, potentially leading to false negative results. The features selected with sequential forward-selection respected the importance values provided by the corresponding regression model. For both, LARS and RF, the first selected feature was  $X_4$ in all cases, then $X_1$ and $X_2$ were included with equal probability. For LARS, the next feature was  $X_5$ in all cases, whereas RF included $X_5$ in about \SI{60}{\percent} and $X_3$ in \SI{30}{\percent} of the runs. If more than four features were to be selected as it is shown in Figure~\ref{fig:friedman_rates_knn}J,  RF always selected the feature set $X_i$ and random additional features from the set $Z_i$ with equal probability. With LARS, however, $X_3$ was not detected as a feature, and was selected randomly and with the same probability as the nuisance variables $Z_i$. Therefore, using forward-selection, LARS failed to include $X_3$ with higher probability than the nuisance variables $Z_i$, while the RF reliably included all relevant variables. However, both methods selected nuisance variables $Z_i$ if a higher number of features was provided as stopping criterion.

\begin{table}[ht]
    \centering
    \begin{tabular}{lrrrrrrrr}
    \toprule
                   & \multicolumn{4}{c}{CMI Criterion} & \multicolumn{4}{c}{MI Criterion} \\
     Model        & TP & TN & FP & FN & TP & TN & FP & FN \\ \midrule
      Friedman I   & \textbf{100.0} &  98.8 & 1.2 &  \textbf{0.0}  & 81.0 & \textbf{99.2} & \textbf{0.8} & 18.4 \\
      Friedman II  &  50.5 &  \textbf{99.8} & \textbf{0.2} & 49.5  & \textbf{56.8} & 99.2 & 0.8 & \textbf{43.3} \\
      Friedman III &   \textbf{3.5} & \textbf{100.0} & \textbf{0.0} & \textbf{96.5}  &  3.3 & 99.5 & 0.5 & 96.8 \\
    \end{tabular}
    \caption{True versus selected features using the proposed forward-selection algorithm with a conditional mutual information (CMI) versus a mutual information (MI) inclusion criterion. Bold numbers indicate the better performing criterion for each measure. Data were simulated according to Friedman models I-III\@. (TP\@: fraction of true positives, TN\@: true negatives, FP\@: false positives, FN\@: false negatives.)}\label{tab:friedman_rates}
\end{table}

Results for models II and III showed lower true positive rates for both information-theoretic criteria~(Table~\ref{tab:friedman_rates}). For model II the algorithm with the CMI criterion failed to select features $X_1$ and $X_4$ (Figure~\ref{fig:friedman_rates_knn}B), for model III almost no features were selected (Figure~\ref{fig:friedman_rates_knn}C, F). Notably, for model II, both criteria successfully selected $X_2$ and $X_3$, while  the number of true positives for $X_1$ identified by the MI criterion was higher than the number for the CMI criterion. However, also the number of false positives increased (Figure~\ref{fig:friedman_rates_knn}E, F).
Over all three models, there was no clear advantage of either the CMI or the MI criterion over the other (Table~\ref{tab:friedman_rates}).

For model II, RF assigned high importance values to the same features as selected using the CMI criterion, $X_2$ and $X_3$, while all other features had negligible importance (Figure~\ref{fig:friedman_rates_knn}H). On the other hand, LARS detected $X_3$ as the only important feature and failed to assign importance to any other feature.
When using forward-selection with LARS, results resembled feature selection performed using the MI criterion, while RF regression always included feature $X_1$ additionally (Figure~\ref{fig:friedman_rates_knn}K). Forward-selection using LARS for model II always selected feature $X_3$ first, and feature $X_2$ was deterministically selected even though its importance value was close to zero. When selecting more than two features, any of the remaining features $X_1$, $X_4$, and $Z_i$ were selected, where $X_1$ was selected consistently more often than the others, which were equally likely to be selected. The RF forward-selection on the other hand, always selected $X_1$ prior to any variable $Z_i$, even though the importance of variable $X_1$ in the full model was as low as the nuisance variables $Z_i$. Features $X_4$ and $X_5$ were included with the same probability as nuisance variables $Z_i$ if the stopping criterion was set to a higher number of features.

For model III, both regression models assigned highest importance to feature $X_3$ and RF regression additionally assigned an importance of around half of the maximum value to features $X_1$ and $X_2$ (Figure~\ref{fig:friedman_rates_knn}I). Using RF regression, Features $X_4$ and $X_5$, and nuisance variables $Z_i$ were assigned an identical importance, around a third of the maximum importance. Using LARS, variables $X_1$, $X_2$ and $X_4$ had coefficients close to zero in all runs. Interestingly, $X_5$ and the nuisance variables $Z_i$, on the other hand, were assigned a small but finite importance by LARS\@. When using forward-selection, both LARS and RF regression selected features $X_1$, $X_2$, and $X_3$ with highest probability, and feature $X_4$ and variables $Z_i$ with identical lower probability (Figure~\ref{fig:friedman_rates_knn}L). Here, the forward-selection with LARS did not follow the order suggested by the importance values. Even though $X_1$ and $X_2$ had much lower importance values, they were reliably selected directly after the dominant variable $X_3$. If more than three variables were allowed to be selected, $X_4$, $X_5$, and the nuisance variables $Z_i$ were included with roughly equal probability.

In sum, compared to CMI and MI, for both models II and III the more traditional methods led to a higher true positive rate for selecting  $X_i$ when a small number of features was used as stopping criterion, but produced substantially higher numbers of false positive for variables $Z_i$ when more than three features were used as stopping criterion.

We evaluated the selected feature sets by testing their performance in predicting the target variable $Y$ using a $k$-nearest neighbor (KNN) predictor with $k=5$ for each Friedman model (Figure~\ref{fig:friedman_rates_knn}M-O). We chose a KNN predictor as it requires little hyperparameter-tuning, except for determining the number of neighbors, $k$. We here report results for $k=5$, while other choices for $k$ led to qualitatively similar results, while only the absolute prediction error changed. We compared the performance of the KNN predictor using the selected feature sets as input against the performance for each possible subset of the complete feature set $\bf{X}$, i.e.\ each element of the power set of $\bf X$, using the mean absolute error (MAE) on \SI{30}{\percent} of samples as test set. For models I and II, the feature set selected by the CMI criterion was the set with the best performance of all possible subsets. While for model I all features were necessary to achieve the best performance, interestingly, for model II the best performance was achieved using the set identified by the CMI criterion which contained only two of the simulated input features, $\{X_2,X_3\}$  (Figure~\ref{fig:friedman_rates_knn}M, N). For model III, we show the predictive performance of all feature sets in the power set, but none of those sets was reliably detected by either criterion. Generally, their performance on model III was rather bad as no set allowed for a prediction with an error comparable to the other two models (Figure~\ref{fig:friedman_rates_knn}O). We conclude that the full feature set may not always be optimal in predicting the dependent variable, as shown for model II, and that for model III a prediction using a simple KNN-model was not successful for any possible feature set.

\begin{figure}[ht]
    \centering
    \includegraphics{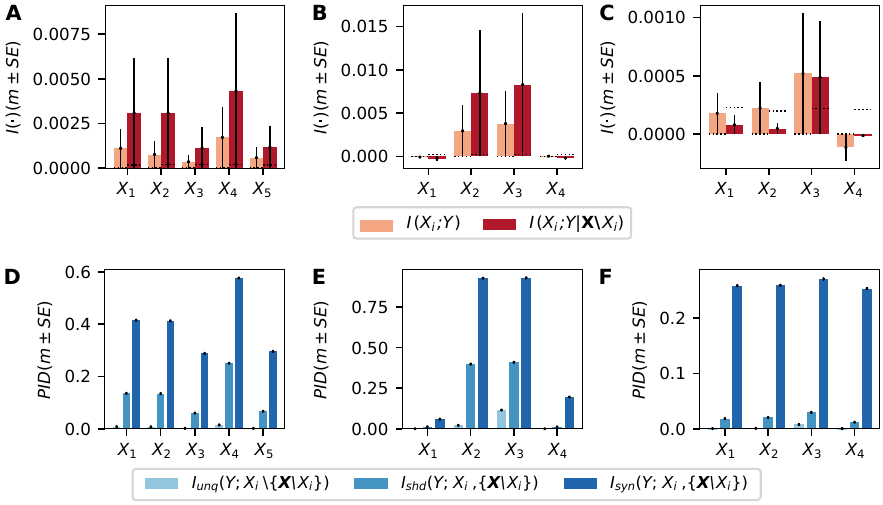}
    \caption{\textbf{Mutual information and partial information decomposition estimates for Friedman models.}
    Mutual information (MI), $I(X_i; Y)$, and conditional mutual information (CMI), $I(X_i; Y|\{\mathbf{X}\setminus X_i\})$, between features, $X_i \in \mathbf{X}$, and target, $Y$, for Friedman model I (\textbf{A}), model II (\textbf{B}), and model III (\textbf{C}) (error bars indicate the standard error of the mean (SEM) for  100 runs). Dashed lines indicate mean significance threshold ($\alpha=0.05$,  $n=200$ permutations).
    Partial information decomposition (PID) atoms between individual features $X_i$, the remaining feature set
    $\mathbf{X}\setminus X_i$, and target $Y$, for Friedman model I (\textbf{D}), model II (\textbf{E}), and model III (\textbf{F}).
    }\label{fig:friedman_validation_pid}
\end{figure}

We estimated the MI, CMI, and PID atoms between individual variables $X_i$, and the target $Y$, and the remaining feature set $\mathbf{X}\setminus X_i$, respectively (Figure~\ref{fig:friedman_validation_pid}). For model I, we found high synergy between each variable and the remaining variable set with respect to the target (Figure~\ref{fig:friedman_validation_pid}D), explaining the higher CMI compared to the MI (Figure~\ref{fig:friedman_validation_pid}A) and accordingly a more successful feature selection when using the CMI criterion. Similar results were observed for model II (Figure~\ref{fig:friedman_validation_pid}B, E). Note that features with high synergy also carried redundant information with the remaining feature set, leading to a higher MI\@. For model III, the total information carried by individual features, either individually or in the context of the remaining feature set, was below the significance threshold for most instances (Figure~\ref{fig:friedman_validation_pid}C). Here, information was almost exclusively synergistic (Figure~\ref{fig:friedman_validation_pid}F), explaining the failure of both forward-selection algorithms in selecting these features.

\subsection{Experiment IV\@: Model by Runge et al. (2015)}\label{sec:experiments_runge}

Next, we generated data using a model proposed by~\cite{Runge2015}, which was designed to generate both synergistic and redundant contributions between time series. For algorithms not accounting for interactions between features, the former property leads to failure to include relevant features, while the latter property leads to the erroneous inclusion of redundant features. The authors show that for this case, a simple forward-selection using a pure CMI-criterion failed (using an algorithm proposed by~\citealt{Tsimpiris2012}).

The model comprises nine input time series, $\mathbf{Z} = \bigcup_{i=1}^3 Z_t^{(i)}$, $\mathbf{W} = \bigcup_{i=1}^4 W_t^{(i)}$, and $\mathbf{X} = \bigcup_{i=1}^2 X_t^{(i)}$, with time indices $t\in\{1,\ldots,T\}$, and one target time series, $Y_t$,
\begin{align*}
    \begin{split}
        Y_{t+1} &= c \sum_{i=1}^4 W^{(i)}_{t-1}
            + b \prod_{i=1}^3 Z^{(i)}_{t-1}
            + \sigma\, \mathcal{N}(0,1), \\ 
        X^{(1)}_t &= a \left( W^{(1)}_{t-1} + W^{(3)}_{t-1} \right)
            + \mathcal{N}(0,1), \\ 
        X^{(2)}_t &= a \left( W^{(2)}_{t-1} + W^{(4)}_{t-1} \right)
            + \mathcal{N}(0,1), 
    \end{split}
\end{align*}
where $\mathbf{W}$ and $\mathbf{Z}$ are i.i.d.\ sampled from Gaussian distributions with zero mean and unit variance ($\sim \mathcal{N}(0,1)$), and the coupling strength between variables is defined by parameters $a = 0.4$, $b = 2$, and $c = 0.4$. The noise in $Y_t$ is scaled by $\sigma = 0.5$.

The three groups of variables, $\mathbf{Z}$, $\mathbf{W}$, and $\mathbf{X}$ are designed to each contribute to $Y_t$ in a different fashion: sets $\mathbf{W}$ and $\mathbf{Z}$ are considered true \enquote{drivers} of the target, $Y_t$, where the sets of variables at time step $t$ determine the target two time steps later,  $Y_{t+2}$. Also, variables in $\mathbf{Z}$ are considered stronger drivers than those in $\mathbf{W}$ due to a higher coupling, $b=2$, compared to $c=0.4$, and $\mathbf{Z}$ is considered to provide mostly synergistic information due to the multiplication of the comprising variables. Set $\mathbf{X}$, on the other hand, is considered to contain redundant variables that carry information already provided by $\mathbf{W}$, as $\mathbf{X}_t$ is determined by the set $\mathbf W$ only one time step earlier, i.e.\ $\mathbf W_{t-1}$. Thus, information observed in $Y_t$ was previously first observed in $\mathbf W_{t-2}$ and then also in $\mathbf X_{t-1}$, which justifies the variables $\mathbf X$ to be considered redundant.

Approaches not considering synergies between variables may fail to include variables in $\mathbf{Z}$. Furthermore, the authors hypothesized that synergistic drivers, $Z^{(i)}$, were individually less predictive than variables $W^{(i)}$, and that variables $Z^{(i)}$ provided information only when considered jointly. Lastly, approaches not accounting for redundancies between variables may erroneously select variables, $X^{(i)}$, additionally to variables, $W^{(i)}$.

\begin{figure}[p]
    \centering
    \includegraphics[width=\textwidth]{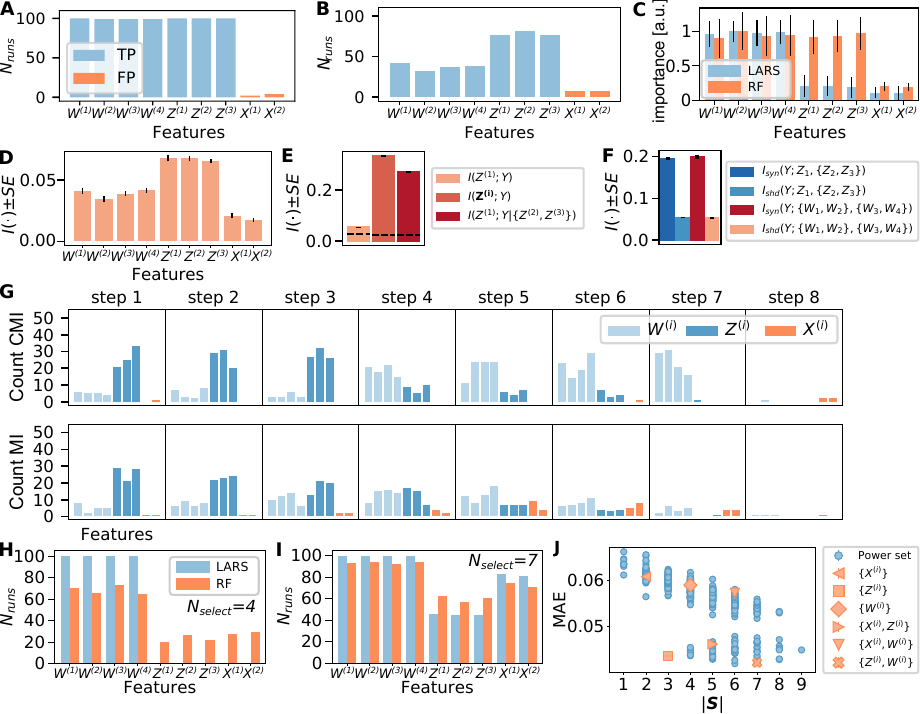}
    \caption{\textbf{Results for model by \protect\cite{Runge2015}.} Number of included variables over 100 simulations using conditional mutual information (CMI, \textbf{A}) and mutual information (MI, \textbf{B}) inclusion criteria and the proposed forward-selection algorithm on data from the model by \protect\cite{Runge2015}. Blue bars indicate correctly included true features (TP\@: true positives), orange bars indicate erroneously included spurious features (FP\@: false positives).
    \textbf{C} Importance values from LARS and RF regression models.
    \textbf{D} MI between individual variables and $Y$.
    \textbf{E} Information contribution of individual variables, $Z^{i}$, versus set $\mathbf{Z}$;
    \textbf{F} Partial information decomposition for variable sets $\mathbf{Z}$ and
    $\mathbf{W}$;
    \textbf{G} Order of inclusion of variables using the CMI and MI criterion.
    Number of included variables using forward-selection with LARS and RF regression models for $N_{select}=4$ (\textbf{H}) and $N_{select}=7$  (\textbf{I}) selected features.
    \textbf{J} Mean absolute error (MAE) of $k=5$-nearest neighbor prediction of $Y$ from all possible subsets of sizes $|\mathbf{S}|$.
    }\label{fig:runge_results_validation}
\end{figure}

We found that the proposed algorithm using the CMI criterion detected true features with great reliability (Figure~\ref{fig:runge_results_validation}A and B, Table~\ref{tab:runge_rates}), while for the MI criterion, false positive and false negative rates increased. In particular, more spurious features, $X^{(i)}$, were selected, while true features were selected less often, in particular weakly predictive variables, $W^{(i)}$.

The feature importance values estimated with the LARS regression provided high values only for the feature set $W^{(i)}$ and assigned the same small importance to the sets $Z^{(i)}$ and $X^{(i)}$ (Figure~\ref{fig:runge_results_validation}C). The RF regression model properly assigned high importance to the features $W^{(i)}$ and $Z^{(i)}$, and very small importance to $X^{(i)}$. However, when performing sequential forward-selection with both models, the selection sequence did not match the ranking of importance values. For example, when allowing for the inclusion of \num{4} features, LARS reliably selected $W^{(i)}$ as features whereas RF selected the $W^{(i)}$ only in about \SI{60}{\percent} of the cases (Figure~\ref{fig:runge_results_validation}H). Furthermore, RF regression correctly included $Z^{(i)}$ in about \SIrange{20}{25}{\percent}, but also wrongly included $X^{(i)}$ in the same number of cases. When allowing for the selection of \num{7} features, both methods correctly selected feature set $W^{(i)}$, but also erroneously selected features $X^{(i)}$ in about \SIrange{70}{80}{\percent} of the runs, while the set $W^{(i)}$ was only selected in about \SIrange{50}{60}{\percent} of the runs (Figure~\ref{fig:runge_results_validation}I). This is in contradiction to the estimation of the features' importance where the feature set $X^{(i)}$ was reliably assigned the lowest importance values. LARS and RF regression thus led to false positive as well as false negative selections as hypothesized by~\cite{Runge2015}.

\begin{table}[ht]
    \centering
    \begin{tabular}{lrrrr}
      \toprule
      Criterion & \multicolumn{1}{c}{TP} & \multicolumn{1}{c}{TN} & \multicolumn{1}{c}{FP} & \multicolumn{1}{c}{FN} \\ \midrule
      CMI       &  \textbf{99.6} &  \textbf{97.0} & \textbf{3.0} &  \textbf{0.4} \\
      MI        &  54.7 &  92.5 & 7.5 & 45.3 \\
    \end{tabular}
    \caption{Actual versus selected features by the conditional mutual information (CMI) versus the mutual information (MI) criterion on data simulated according to \protect\cite{Runge2015}. Bold numbers indicate the better performing criterion for each measure.}\label{tab:runge_rates}
\end{table}

Results obtained from the proposed information-theoretic algorithm partially contradict the hypothesized behavior of forward-selection algorithms on the generated data, for example, the hypothesis that simple MI selection criteria should fail to identify features in $\mathbf{Z}$, because they do not account for the assumed synergistic information contribution. Here, further investigation of the simulated data showed that variables in $\mathbf{Z}$ provided individually more information about $Y$ than variables in $\mathbf{W}$ and $\mathbf{X}$, with $\mathbf{X}$ providing the least information (Figure~\ref{fig:runge_results_validation}D). We found that already individual variables, $Z^{(i)}$, provided significant information about $Y$, $I(Y;Z^{(i)})$ (Figure~\ref{fig:runge_results_validation}E). This contribution increased if the whole set $\mathbf{Z}$ was considered by estimating $I(Z^{(1)};Y_t|\{Z^{(2)}, Z^{(3)}\})$ (Figure~\ref{fig:runge_results_validation}E). Also, the synergistic contribution by $\mathbf{Z}$ was comparable to the synergistic contribution of $\mathbf{W}$ (Figure~\ref{fig:runge_results_validation}F). Here, the multiplication of variables did not introduce a stronger synergistic effect compared to the additive combination of variables (see also Experiment II and~\citealt{Griffith2014}).

Furthermore, the authors hypothesized that spurious drivers, $X^{(i)}$, should individually provide more information about $Y_t$ than variables $W^{(i)}$ and $Z^{(i)}$, and that this should be reflected in the order of inclusion during forward-selection, which was expected to be $X^{(i)} \rightarrow Z^{(i)} \rightarrow W^{(i)}$ for a CMI criterion, and $X^{(i)} \rightarrow W^{(i)} \rightarrow Z^{(i)}$ for an MI criterion. However, we found that the order of inclusion in our algorithm followed the order of the magnitude of individual information contribution, $Z^{(i)} \rightarrow W^{(i)} \rightarrow X^{(i)}$ (Figure~\ref{fig:runge_results_validation}G), as determined by the coupling strength $b=2>c=0.4$. Accordingly, in the first three inclusion steps, variables $Z^{(i)}$ were predominantly included by both the MI and CMI criterion, followed by variables $W^{(i)}$ in the next steps. From step 4 on, the CMI identified more variables from $W^{(i)}$ while the MI criterion failed to include these variables and instead included more spurious variables, $X^{(i)}$.

Finally, we show the performance of the various selected feature sets in the task of predicting the output variable $Y$ using a $k=5$ KNN predictor. Figure~\ref{fig:runge_results_validation}J shows the MAE of the KNN predictors, each predictor trained with one element of the power set of the complete feature set as input, and with \SI{30}{\percent} of the samples as a test set. Some special subsets of features are marked as shown in the legend. It can be observed that indeed the predictor using only feature set  $Z^{(i)}$, which are the three features selected during the first three iterations using the MI and CMI criterion, gave the best mean performance. In particular, this predictor provided the best trade-off in terms of minimum number of input features and prediction performance, and its error was significantly lower than with all other features sets of size $|S|=3$. Inclusion of the feature sets  $Z^{(i)}$  and $W^{(i)}$, as suggested by the CMI criterion and partly by the MI criterion and the importance values of the RF, resulted in a comparable prediction performance. In contrast, the feature set $W^{(i)}$, which was selected first by the LARS (importance values and sequential forward-selection) and RF (sequential forward-selection) approaches, led to substantially worse prediction performance. Extending the features set by including $W^{(i)}$  and $X^{(i)}$, as suggested by sequential forward-selection with LARS and RF, resulted in an equally bad prediction.

\subsection{Experiment V\@: Noise Examples by MacKay (2005) and Haufe et al. (2014)}

Synergistic effects do not only occur in purely artificial examples, but also appear in simple real-world settings such as noisy measurements. We analyzed two examples from literature that illustrate the effect of recording noise during data collection \citep{MacKay2005,Haufe2014}. Both examples assume two input variables, $X_1$ and $X_2$, where variable $X_1$ is the measurement of a signal of interest, $S$, that affects the output $Y$ by some function $f(\cdot)$, while the second variable, $X_2$, measures a \enquote{distractor signal}, $D$, that is considered to be noise. A weighting factor, $a \in [0, 4]$, determines the strength of the distortion. Both $S$ and $D$ are drawn from a standard normal distribution. We simulated $f(S)$ as a noisy sinus, i.e., $f(S) = \sin(S) + \sigma\, \mathcal{N}(0,1)$, with $\sigma=0.1$.

\begin{figure}[ht]
    \centering
    \includegraphics[width=0.99\linewidth]{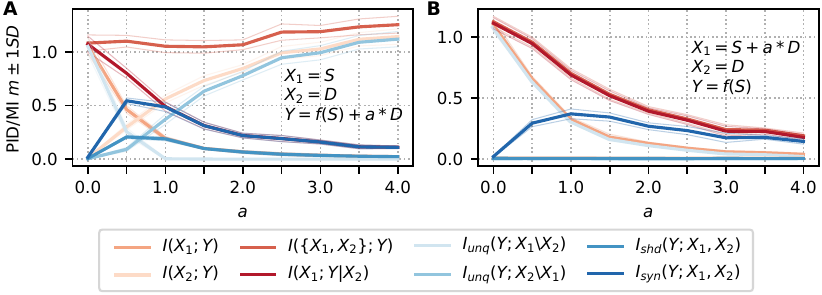}
    \caption{
        \textbf{Results noise examples.} Mutual information (MI) and partial information decomposition (PID) terms for noise examples.
        \textbf{A} Model as described in \protect\cite{MacKay2005}.
        \textbf{B} Model as described in \protect\cite{Haufe2014}. Note that the joint MI, $I(Y;\{X_1,X_2\})$, is covered by the CMI, $I(Y;X_1|X_2)$, and the MI, $I(Y;X_1)$, is covered by the unique information, $I_{unq}(Y;X_1 \setminus X_2)$.
    }\label{fig:noise_models_pid}
\end{figure}

In the first model, formulated after~\cite{MacKay2005}, we consider the measurement of the target variable to be corrupted by the distractor signal,
\begin{equation*}
    \begin{aligned}
        X_1 &= S \,, X_2 = D \\
        Y &= f(S) + aD, \\
    \end{aligned}
    \label{eq:mackay_noise}
\end{equation*}
while in the second model \citep{Haufe2014}, we consider a noisy measurement of the input variable,
\begin{equation*}
    \begin{aligned}
        X_1 = S +& aD \,, X_2 = D \\
        Y &= f(S). \\
    \end{aligned}
    \label{eq:haufe_noise}
\end{equation*}
For the MacKay example \citep{MacKay2005}, we found that the CMI, $I(X_1;Y|X_2)$, at all values of $a$ captured the unique influence of the signal of interest, $X_1$, on $Y$, and included the synergistic contribution of both $X_1$ and $X_2$ (Figure~\ref{fig:noise_models_pid}A). Simultaneously, the unique influence of the noisy signal, $X_2$, and the redundant information in $X_1$ and $X_2$ was conditioned out. The MI, as expected, failed to capture the relationship between the three measurements, where the information provided by $X_1$ alone, $I(X_1;Y)$, decreased for $a \geq 0.5$ and became smaller than the CMI, $I(X_1;Y|X_2)$. The MI thus failed to capture all information provided by the signal of interest $X_1$, which is partially \enquote{masked} as a synergistic contribution between $X_1$ and $X_2$. On the other hand, the information provided by $X_2$ alone, $I(X_2;Y)$, increased and for $a \geq 1.0$ became larger than the CMI, $I(X_1;Y|X_2)$, thus over-estimating the influence of the noisy signal on $Y$. Due to the addition, shared and synergistic information contribution decreased similarly for larger $a$, while the synergistic contribution was always higher than the redundancy between both variables. Lastly, the multivariate MI, $I(\{X_1, X_2\};Y)$, included both the information of $X_1$ and the noisy signal $X_2$, thus failing to decompose the distractor signal into its synergistic contribution required to decode the signal of interest, $X_1$, and its distorting unique and redundant contribution.

For the second example by~\cite{Haufe2014}, we found again that the CMI, $I(X_1;Y|X_2)$, successfully captured the information provided by both variables, while the MI between each variable and $Y$ failed to account for the synergistic contribution of both variables (Figure~\ref{fig:noise_models_pid}B). In this example, also the multivariate MI, $I(\{X_1, X_2\};Y)$, was able to capture the joint contribution of both measurements due to an absence of redundant information between both variables. Here the contribution is purely synergistic, such that no redundancies have to be conditioned out from the measurement.

Both examples show that we can recover a signal of interest also in the presence of high noise if we are able to record the source of the noise. In this case, the recorded noise, $X_2$, may be used to \enquote{decode} the information the signal of interest, $X_1$, carries about the target, $Y$. As already stated by~\cite{Haufe2014}, this is important in the interpretation of subsequent modeling. The relevancy of a variable may be indirect such that the weight attributed to the variable in a model may not encode its immediate relevancy to the target quantity but may indicate its strength in moderating the relationship between the input and output signal. An application example may be the recording of physiological signals \citep{Haufe2014}, where the signal of interest is electrophysiological brain activity, while the distractor signal is the electrocardiogram (ECG). The target variable may be a further brain signal, but also behavioral measurements, e.g., the success in a task. To model the relationship between the electrophysiological recording and the target quantity, accounting for the noise generated by the ECG is beneficial in both scenarios of noise contamination.

\subsection{Experiment VI\@: Examples by Guyon \& Elisseeff (2003) and Haufe et al. (2014)}

As a last example, we analyzed data from two models that investigate whether correlation (more generally, dependency) calculated exclusively between input variables indicates redundancy or synergistic contributions with respect to a target variable, following toy examples in~\cite{Guyon2003} and~\cite{Haufe2014}. It has been stated that the goal of feature selection is the identification of \enquote{independent} features such as to include features that provide a maximum of novel information about the target variable and to avoid redundant features. Accordingly, some approaches use the MI between features as a proxy for redundant information \textit{about} the target. However,~\cite{Guyon2003} previously demonstrated on toy systems that correlation does not indicate feature redundancy and that, on the other hand, the absence of correlation does not indicate an absence of interaction, for example, in the form of synergistic contributions. We here illustrate this by estimating MI, CMI, and PID from similar toy models, where we show that statistical dependence or independence \textit{between} features is not sufficient to identify feature interactions that render features relevant or irrelevant \textit{with respect to the target}.

For all examples, data were independently sampled from two two-dimensional normal distributions with means $\bm{\mu}_1$ and $\bm{\mu}_2$ and common covariance matrices, $\Sigma_{\mathbf {X}, \mathbf {X}}$. Class labels, $Y\in\{0, 1\}$, indicate from which distribution a sample was drawn. As input variables, we considered the two dimensions, $X_1$ and $X_2$.

\begin{figure} 
    \centering
    \includegraphics[width=0.87\textwidth]{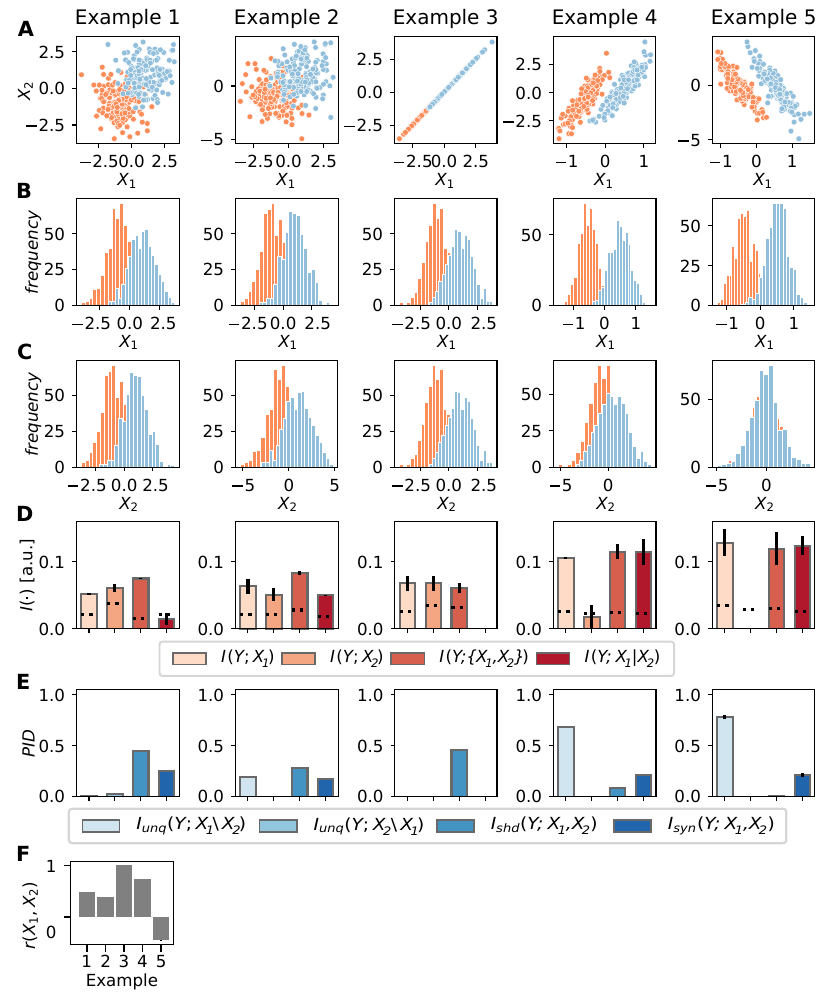}
    \caption{\textbf{Results for models from \protect\cite{Guyon2003}.}
    \textbf{A} Data from exemplary run for Examples 1 to 5 (left to right),
    color indicates class label.
    \textbf{B} Class-conditional distributions of variable $X_1$.
    \textbf{C} Class-conditional distributions of variable $X_2$.
    \textbf{D} Estimated mutual information (MI), conditional mutual information (CMI) and joint mutual information (average over 20 runs, dashed lines indicate mean significance threshold, error bars indicate the standard error of the mean).
    \textbf{E} Estimated partial information decomposition (PID) atoms.
    \textbf{F} Pearson correlation between input variables.
    }\label{fig:guyon2003_results}
\end{figure}

Examples 1 and 2 illustrate that two seemingly non-interacting variables may provide synergistic information. We simulate two sets of independent variables with $\bm{\mu}_1=(-1,-1)^T$ and $\bm{\mu}_2=(1, 1)^T$, respectively. The examples vary in their covariance matrices to illustrate that whether variables may provide synergistic information is dependent on their variance,
\begin{equation*}
    \Sigma^1_{\mathbf {X}, \mathbf {X}}=
    \begin{pmatrix}
        1 & 0 \\
        0 & 1 \\
    \end{pmatrix},
    \Sigma^2_{\mathbf {X}, \mathbf {X}}=
        \begin{pmatrix}
            1 & 0 \\
            0 & 2 \\
        \end{pmatrix}.
\end{equation*}
Example 3 illustrates that only if variables are perfectly correlated, they exclusively provide redundant information. For this we sample the two distributions with means $\bm{\mu}_1=(0, 0)^T$ and $\bm{\mu}_2=(0,0)^T$, respectively, and
\begin{equation*}
    \Sigma^3_{\mathbf {X}, \mathbf {X}}=
    \begin{pmatrix}
        1 & 1 \\
        1 & 1 \\
    \end{pmatrix}.
\end{equation*}
Example 4 illustrates that high correlation does not imply an absence of synergistic information. We simulate the two sets of variables with means $\bm{\mu}_1=(-0.5, 0.5)^T$ and $\bm{\mu}_2=(0.5, 0.5)^T$, and
\begin{equation*}
    \Sigma^4_{\mathbf {X}, \mathbf {X}}=
    \begin{pmatrix}
        0.1 & 0.4 \\
        0.4 & 2 \\
    \end{pmatrix}.
\end{equation*}
Example 5 illustrates that high correlation does not imply an absence of synergistic information and that even a variable that does not provide any information by itself may provide synergistic information with another feature. Means are set to $\bm{\mu}_1=(-0.5, 0)^T$ and $\bm{\mu}_2=(0.5, 0)^T$, and
\begin{equation*}
    \Sigma^5_{\mathbf {X}, \mathbf {X}}=
    \begin{pmatrix}
        0.1 & -0.4 \\
        -0.4 & 2 \\
    \end{pmatrix}.
\end{equation*}
Figure~\ref{fig:guyon2003_results} shows representative samples from all five example distributions (Figure~\ref{fig:guyon2003_results}A-C), as well as estimated information-theoretic quantities (Figure~\ref{fig:guyon2003_results}D-E), and the Pearson correlation coefficient between the inputs for each example (Figure~\ref{fig:guyon2003_results}F). Examples 1 and 2 illustrate that seemingly independent variables may provide synergistic information. However, only in Example 2, where the variance of $X_2$ was higher, a significant information contribution as measured by the CMI criterion was consistently found. In Example 1, the added unique and synergistic contribution reached statistical significance for only some instances. Here, the higher relative redundancy may have rendered $X_1$ not informative given $X_2$ as measured by the CMI criterion. While the addition of $X_1$ may add some improvement in a subsequent learning task, its inclusion also adds a high amount of redundant information and may have to be balanced against the minimization of the feature set size. In both Examples 1 and 2, the pairwise as well as the joint MI was significant in most cases. We conclude that---in line with~\cite{Guyon2003}---\enquote{presumably independent variables} may contribute information through a synergistic interaction, which becomes particularly evident in Example 2.

For Example 3, as hypothesized for perfectly correlated input variables, we found exclusively redundant information in the two input variables, with respect to the target, and no unique or synergistic information contribution. Accordingly, the CMI, $I(Y;X_1|X_2)$, was zero as $X_1$ provided no information about $Y$ if $X_2$ was already known.

Lastly, for Examples 4 and 5, we found in both cases that despite a high correlation between the inputs, there was little redundancy and even some synergistic information contribution between both inputs. The main contribution was provided uniquely by $X_1$. Note that even though the correlation between variables in Example 4 was higher than for Examples 1 and 2, the redundancy was lower.

In sum, through the estimation of the PID, we showed that presumably independent variables may still provide synergistic information about the target, while a dependence between variables is not necessarily indicative of variable redundancy or the absence of synergy. We conclude that measuring the dependence between features is not a suitable proxy to infer feature interactions in general. Our PID-based results are in accordance with the observations by~\cite{Guyon2003}, providing a quantitative description of the relationship between variable dependence and relevance.

\section{Discussion}
We use the partial information decomposition (PID) framework by~\cite{Williams2010} to provide a rigorous, information-theoretic definition of relevancy and redundancy in feature-selection problems that, in particular, accounts for feature interactions and the resulting redundant and synergistic information contributions about a target. We show that classical information theory lacks methods to describe such detailed information contributions and that only through the introduction of the PID framework these methods have become available. Based on our definition of feature relevancy and redundancy in PID terms, we show that using the conditional mutual information (CMI) as a feature selection criterion, e.g., in sequential forward-selection, simultaneously maximizes a feature set's relevancy while minimizing redundancy between features. For practical application, we propose the use of a recently introduced CMI-based forward-selection algorithm that uses a novel statistical testing procedure to handle bias when estimating information-theoretic quantities from finite data. By applying the proposed algorithm to benchmark systems from literature, we demonstrate that the CMI is in almost all cases the preferable selection criterion, compared to the mutual information (MI) which does not account for feature interactions. Additionally, we demonstrate how PID allows for a detailed, quantitative description of individual and joint feature contributions---in particular, for interacting features---that was not possible before using information-theoretic methods.

In the following we discuss relevant aspects of the proposed methodology in more detail.

\subsection{PID Enables the Detailed Quantification of Joint Information Contribution for Interacting Features}

Basing the definition of variable relevancy and redundancy on the PID framework provides a first, rigorous definition of these concepts in information-theoretic terms and makes individual contributions quantifiable using estimators for PID\@. Our definition further resolves existing issues in the description of multivariate information contribution such as negative information contribution and allows making conjectures about variable contributions testable (as formulated, for example, by~\citealt{Guyon2003} or~\citealt{Brown2012}). As demonstrated, similar insights into the \enquote{structure} of the multivariate information contribution in feature selection can not be obtained from using classical information-theoretic terms, but require the introduction of additional axioms as done by the PID framework.

By applying the PID framework to benchmark data sets from literature, we tested a series of earlier conjectures and assumptions about feature interactions. We showed that multiplicative combinations of terms lead to lower synergistic contribution than additive ones, opposed to assumptions made in earlier work. We further showed that determining the dependency \textit{between} features is not sufficient to determine feature relevancy or redundancy with respect to the target variable (as also discussed, for example, by~\citealt{Guyon2003}). In particular, we showed that the absence of correlation did not indicate a lack in synergistic contribution and thus a lack in relevancy. Furthermore, we showed that the presence of correlation did not indicate feature redundancy. Instead of using variable dependency as a proxy to infer such interactions, PID allowed us to immediately quantify these individual and joint contributions \textit{about} the target. Furthermore, we show that both synergistic and redundant contributions can coexist already in simple feature selection problems, such that whether a variable is included in a feature set, hinges on whether its contribution of \textit{novel} information outweighs the information it redundantly shares with other features. Also, we demonstrate that all three unique, synergistic, and redundant contributions can occur in the data simultaneously and with arbitrary shares.

An important limitation of using PID to understand feature selection problems is the number of variables that can be considered simultaneously. When looking at fine-grained interactions between increasingly larger data sets, the number of possible interactions soon becomes intractable as also higher-order interactions between subsets become possible. For example, when considering four input variables that interact with a single target variable, there are already 166 possible ways of how subsets of variables can multivariately provide information about the target \citep{Makkeh2021}. Instead, PID can always be used to quantify the information contribution of a single feature with respect to the \textit{set} of all remaining selected features, as done in the present work. Here, however, the dimensionality of the estimation space is a further limiting factor, restricting the application of existing PID estimators to feature sets of relatively small size.

\subsection{A Remark on PID for More Than Two Input Variables}

We note that we have restricted our application of PID here to the case of two input variables---where one variable represented a particular feature while the other represented the set of all other available features. Our self-restriction was motivated by our focus on the iterative inclusion scenario, which is most relevant practically. However, PID can also be defined for more than two input variables that carry information about an output variable as shown in Appendix~A \citep{Williams2010,Makkeh2021,Gutknecht2021}. In that case we have to ask questions of a more complex type, such as: \enquote{How much of the information that a set of variables $\{X_1, X_2\}$ carries jointly (i.e., also synergistically) about a target variable $Y$ is also carried redundantly with the information that another set of variables, say $\{X_3, X_4, X_5\}$ carries about $Y$?}. The structure of the underlying decomposition can be derived solely relying on what it means for one thing to be part of another (i.e.,~parthood relations or mereology) and is isomorphic to certain well-known lattices, such as the lattice of antichains and lattices from Boolean logic (see~\citealt{Gutknecht2021} for an accessible introduction).

For limited data, the fine-grained decomposition provided by a fully multivariate information decomposition may be impractical, as the number of parts or \enquote{atoms} of the decomposition scales extremely fast with the number of input variables (here, features): for $N$ input variables the number of parts is equal to $D(N)-2$, where $D(N)$ is the $N$-th Dedekind number. For example, for $N=2,3,4,5,\ldots$ inputs the corresponding number of parts are $4, 18, 166, 7579,\ldots$, respectively. While this rapid scaling is frustrating from a practical point of view, it at least offers a clean quantitative measure of the inherent (and unavoidable) complexity of the feature selection problem; in other words, the problem is certainly \NP-hard, but still the Dedekind numbers rise much slower than the cardinality of the power set of all features, which would have to be considered when approaching the problem naively.

\subsection{Relationship to Definitions of \enquote{Strong and Weak Relevancy}}

In the context of information-theoretic feature selection, feature relevancy has been mathematically defined in probabilistic as
well as information-theoretic terms (e.g.,~\citealt{John1994,Bell2000}, see also~\citealt{Vergara2014} for a review). It has been recognized that common \enquote{definitions give unexpected results, and that the dichotomy of relevancy vs irrelevance is not enough} \citep{Bell2000}. Therefore,~\cite{John1994} introduced the concept of \emph{strong and weak relevancy} to mitigate these problems, in particular to avoid results contradicting the intuitive notion of relevancy in the presence of fully redundant features. The definition by John et al.\ can be written in information-theoretic terms \citep{Vergara2014}, such that for a set of input variables $\mathbf{X}$ and a dependent variable $Y$, a feature, $F$, from the set of all true features, $\mathbf{S}$ is defined as \textit{strongly relevant} if
\begin{equation}
    I(F; Y | \mathbf{S} \setminus F) > 0,
    \label{eq:strongly_relevant}
\end{equation}
where $\mathbf{S} \setminus F$ the set of selected features excluding $F$. A strongly relevant feature provides information about $Y$ that can not be obtained from any other feature. From a PID perspective, we can see that this criterion ensures that $F$ provides either unique or synergistic information, or both, which leads to a positive CMI\@.

Next, a variable is defined to be \textit{weakly relevant} if
\begin{equation*}
    \begin{aligned}
        I(F; Y | \mathbf{S} \setminus F) &= 0 \;\; \land \\
        \exists \mathbf{S}' &\subseteq \mathbf{S}\setminus F:\,  I(F; Y | \mathbf{S}') > 0
        \label{eq:weakly_relevant}
    \end{aligned}
\end{equation*}
i.e., the feature is not strongly relevant and there exists a subset of features, $\mathbf{S}'$, in whose context $F$ provides information about $Y$. Thus, weakly relevant features \enquote{can sometimes contribute} information in the context of a suitably selected subset, $\mathbf{S}'$ \citep{Bell2000}. Again, using PID we can state that a weakly relevant feature provides information that is fully redundant with the full feature set, or more precisely, with a subset of the feature set, $\mathbf{S}\setminus \mathbf{S}'$. Note that all variables that are perfect copies of another feature are fully redundant with respect to $Y$ and provide neither unique nor synergistic information. On the other hand, not all features that provide purely redundant information about a target are necessarily perfect copies of each other. It is also conceivable that two variables provide completely redundant information about the target and carry further information that is independent of the target.

Lastly, a feature is considered irrelevant if
\begin{equation*}
    I(F;Y|\mathbf{S}') = 0 \;\; \forall \, \mathbf{S}' \subseteq \mathbf{S} \setminus F,
    \label{eq:irrelevant}
\end{equation*}
i.e., the feature never provides information about $Y$ and can be discarded. Again, from a PID point of view, a feature is not included if it provides information neither uniquely of synergistically in the context of the remaining feature set and all possible subsets, $\mathbf{S}'$.

Our definition of feature relevancy and the proposed forward-selection algorithm using a CMI criterion are in line with the definitions by~\cite{John1994}, which requires for a suitable feature selection algorithm to select all strongly relevant features, the smallest possible subset of weakly relevant features, while not including any irrelevant features \citep{John1994}. Note further that the definition of weak and strong relevancy is not directly applicable in practice because it requires the evaluation of the power set of $\mathbf{S}$. Last, it should be noted that when investigating strongly and weakly relevant features using PID measures, the choice of the measure  is important. In particular, the measure should be sensitive to \enquote{what} information over \enquote{how} much information is provided, which is not the case for all proposed PID measures (see for example,~\citealt{Harder2013,Bertschinger2014,Timme2018}). Measures from the latter category may lead to counter-intuitive results, while the BROJA measure used here does not suffer from this problem. Despite the remaining issues in its estimation, we conclude that the PID may provide a more straightforward and more intuitive characterization of information contribution than weak and strong relevancy.

\subsection{Relationship to Feature-Selection Framework by Brown et al. (2012)}

The framework development by~\cite{Brown2012}  subsumes many feature selection criteria under a common formulation. The authors arrive at this formulation by first showing that the CMI is the optimal criterion for variable inclusion in sequential forward-selection, when considering filtering features prior to fitting a classifier as the problem of maximizing the conditional likelihood of the correct class labels given the selected features. After showing the optimality of the CMI, the authors show that most existing inclusion criteria can be related to the CMI under a series of independence assumptions, where independence assumptions are typically made to avoid the estimation of the full CMI in potentially high-dimensional feature spaces.

In particular, the framework first assumes independence and class-conditional independence for all non-selected variables $X_k \in \mathbf{X}_i$ at inclusion step $i$, i.e., for all $X_k \in \mathbf{X}_i$,
\begin{equation}
    \begin{aligned}
        p(\mathbf{S}_i|X_k) &= \prod_{F_j \in \mathbf{S}_i} p(F_j|X_k), \\
        p(\mathbf{S}_i|X_k,Y) &= \prod_{F_j \in \mathbf{S}_i} p(F_j|X_k,Y).
        \label{eq:assumption_1}
    \end{aligned}
\end{equation}

The CMI criterion in Equation~\eqref{eq:cmi_criterion} can be rewritten as (see~\cite{Brown2012} for a detailed derivation),
\begin{equation*}
    I(Y; F_i|\mathbf{S_i}) = I(Y; F_i)
            - \sum_{F_j \in \mathbf{S}_i} I(F_i; F_j)
            + \sum_{F_j \in \mathbf{S}_i} I(F_i; F_j|Y).
\end{equation*}
Here, the second term can be discarded under the assumption of pairwise independent variables, i.e., $p(F_i; F_j) = p(F_i)p(F_j)$. The third term can be discarded under the assumption that all features are pairwise class-conditionally independent, i.e., $p(F_i; F_j|Y) = p(F_i|Y)p(F_j|Y)$. Furthermore, both terms can enter the equation to varying degrees, expressing a corresponding degree of belief in each assumption, yielding the formulation
\begin{equation}
    J^\prime_{CMI}(F_i) = I(Y; F_i)
            - \beta \sum_{F_j \in \mathbf{S}_i} I(F_i; F_j)
            + \gamma \sum_{F_j \in \mathbf{S}_i} I(F_i; F_j|Y),
    \label{eq:brown_framework}
\end{equation}
which describes a generic inclusion criterion, $J^\prime_{CMI}$, derived from the CMI under the three assumptions above. Here, lower values for either parameter lessens the contribution of the respective term, indicating that the corresponding independence-assumption is adopted to a higher degree. Many existing information-theoretic criteria may be expressed by Equation~\eqref{eq:brown_framework} through the adoption of specific values for $\gamma$ and $\beta$ (see~\citealt{Brown2009}).

Several observations can be made at this point: First, the assumption made in Equation~\eqref{eq:assumption_1} discards all higher-order interactions between features. Hence, all criteria that can be subsumed under this framework only account for pairwise interactions between features.

Second, the second term in Equation~\eqref{eq:brown_framework} ($\sim\beta$) leads to a decrease in a feature's relevancy whenever that feature has high MI about any feature already selected. Thus, these criteria use the pairwise MI between the feature under consideration and already selected features to quantify feature redundancy, an assumption that---as already discussed---may not generally hold. Moreover, the MI between $F_i$ and $F_j$ is in general independent of the redundancy the two variables carry about the target, such that positive MI does not necessarily indicate redundant information in the features about the target, nor does an MI of zero indicate an absence of redundant information. The first statement may be illustrated with an example, where each feature, $F_i$ and $F_j$, is a combination of two further variables, $F_i=\left\{X_1, X_2\right\}$, $F_j=\left\{Z_1, Z_2\right\}$, which are not observed and thus not part of the input variable set. Here, $I(F_i;F_j) > 0$ may be due to $I(X_1;Z_1)>0$, while the information $F_i$ and $F_j$ provide about $Y$ is due to $I(X_2;Y)>0$ and $I(Z_2;Y)>0$, while $I_{shd}(Y;\{F_i,F_j\})=0$. In this scenario, the MI, $I(F_i;F_j)$, is non-zero while the information provided by both variables about $Y$ is not redundant. The second statement, i.e., zero MI does not indicate the absence of redundancy, may be illustrated by a simple logical AND gate with two independent, Boolean inputs, where the BROJA measure used here returns $I_{shd}=0.311$ \citep{Bertschinger2014}. This type of redundant information contribution has also been termed \textit{mechanistic redundancy}, i.e., redundancy due to the operation performed on the input to generate the output, and has to be distinguished from redundancy that arises from information redundantly present in the input, termed \textit{source redundancy} \citep{Harder2013}. In the special case, where just source redundancy occurs in a system, the MI between the inputs is an upper bound on the redundancy if the PID measure used adheres to the identity axiom introduced by~\cite{Harder2013}.

Third, criteria not including the third term in Equation~\eqref{eq:brown_framework} (i.e.\ when $\gamma=0$) do not account for pairwise synergistic information between $F_i$ and $F_j$ with respect to $Y$. Hence, if this term is not included, the criterion may be overly conservative because it misses a feature's synergistic relevancy, which---as demonstrated in the experiments---is present already in simple systems.

\subsection{CMI as an Optimal Feature-Selection Criterion and Estimation from Data}

We demonstrated that already in simple examples, feature interactions can lead to relevant synergistic contributions or redundancy. Accordingly, the CMI criterion, which accounts for such interactions, outperformed the MI criterion in almost all investigated scenarios. Our results therefore highlight the importance of developing feature selection criteria that account for interactions between variables; assuming variable independence may easily yield non-optimal solutions that both miss relevant variables and falsely include redundant ones. Also for practical applications, it has been noted that already simple problems show interactions between features with respect to the target such that considering pairwise dependencies alone leads to an insufficient characterization of the system under investigation \citep{Reing2018}. Examples include data sets obtained from sufficiently complex, multivariate real-world systems, such as biological or artificial neural networks \citep{Latham2005synergy,Tax2017}, or ecological models \citep{Grilli2017}.

Yet, existing inclusion criteria for sequential forward-selection (e.g.,~\citealt{Duch2006,Brown2012}) often fail to account for features interactions, in particular, for synergistic effects (e.g.,~\citealt{Cheng2011,Vergara2014,Shishkin2016}). Instead, many criteria make the (implicit) modeling assumption that variables act on the target independently \citep{Dash1997,Hall2000}, or that variables interact only pairwise with respect to the target \citep{Brown2012}. These approaches aim at avoiding (C)MI estimation in high-dimensional spaces \citep{Brown2012,Vergara2014}. Also, approaches explicitly accounting for synergistic contributions often use low-dimensional approximations of the synergy that may miss relevant contributions \citep{Zhao2007,Tsanas2010,Cheng2011}. Here, PID may help to construct novel approaches as it allows to characterize also intermediate levels of dependency, thus covering the middle ground between pairwise and the total dependency in a set of variables \citep{Reing2018}.

For medium-sized feature selection problems, estimators with more favorable bias properties may be used if applicable. We here use a continuous nearest-neighbor based estimator that has been shown to perform well in relatively high-dimensional spaces \citep{Kraskov2004}, and that has favorable bias properties \citep{Kraskov2004,Khan2007,Lizier2014jidt,Xiong2017}, in particular compared to plug-in estimators for discrete or discretized variables (e.g.,~\citealt{Doquire2012}), and has been successfully applied in feature selection \citep{Francois2007,Tsimpiris2012,Doquire2012,Lensen2018}. However, even though the estimator shows better bias properties than comparable approaches, it is not bias-free and still requires the statistical testing of estimates against suitable surrogate data (e.g.,~\citealt{Francois2007}). We here propose to handle estimation bias using a recently proposed statistical testing scheme that---opposed to comparable approaches (e.g.,\citealt{Francois2007,Tsimpiris2012})---explicitly controls the false-positive rate during feature selection \citep{Novelli2019,Wollstadt2019}. The testing scheme thereby ensures that feature inclusion terminates when the remaining variable set contains no more relevant features, but also if a potential feature's contribution can no longer be robustly estimated given the dimensionality of the already selected feature set and the available data. While this approach avoids false positives, it may be overly conservative in cases where the number of relevant features is high relative to the amount of available data. Such cases may be handled by considering only subsets of the already selected feature set when estimating the CMI \citep{Fleuret2004}.

Finally, we want to note that forward-selection algorithms, such as one the used here, may fail to include feature (sub-)sets with purely synergistic contributions (as shown in Experiment I, Section~\ref{sec:experiment_spheres}, and Experiment III, Section~\ref{sec:experiments_friedman}). One may mitigate this problem by initialize the feature set not as an empty set but with random subsets of variables, or one may try to include tuples of increasing size based on their joint information contribution \citep{Lizier2012}. However, these approaches are limited to problems with few input variables due to their practical run time.

\subsection{Conclusion and Outlook}

In the present paper, we presented a rigorous definition of feature relevancy and redundancy based on the recently proposed PID framework. We introduced PID as a conceptual as well as a practical framework for the investigation of information contribution in feature selection problems, in particular, for the analysis of feature interactions that give rise to redundant and synergistic information contributions. Based on our definition of feature relevancy, we showed that the CMI is the optimal criterion in sequential feature selection that simultaneously accounts for synergistic information contribution, while avoiding redundancy between features. We presented a forward-selection algorithm for the practical application of the CMI as a selection criterion, which handles estimator bias and provides a robust stopping criterion for variable inclusion. We successfully applied the algorithm and PID to illustrative examples from literature, illustrating how PID leads to a better understanding of how features and their interactions contribute information in feature-selection problems. Future work may extend the use of PID in feature selection, in particular, in the definition of inclusion criteria that properly account for feature interactions.

\newpage
\appendix

\section{Multivariate Partial Information Decomposition}\label{app:pid_lattice}

Using the partial information decomposition (PID) framework \citep{Williams2010}, we showed in the main text that the conditional mutual information (CMI) is the better choice for a criterion to select features than the mutual information (MI) for the case of two input variables or sets of variables. The argument can be extended to the general, multivariate case as shown in the following. The structure of a full PID for a multivariate MI, $I(Y;\{F_1,\ldots,F_n\})$, can be derived based on the insight that each information atom can be characterized by its distinctive parthood relationships to the information provided by the different possible collections of features (e.g.,~the highest-order synergy being part of the information provided by the collection of all features together but not part of the information provided by any subcollection). Starting from a small set of basic assumptions about parthood relationships between information contributions one may infer that there is a definite number of different information atoms for a given number of features and that these atoms can naturally be ordered into a mathematical structure, alternatively called a parthood lattice, the lattice of antichains, or the redundancy-lattice \citep{Gutknecht2021}.
The redundancy lattice for PID describes a partially ordered set in which each element contains at least as much redundant information as the lower ones, and which allows to describe the decomposition of a joint MI into its PID terms.

\begin{figure}[p]
    \centering
    \includegraphics[width=0.85\textwidth]{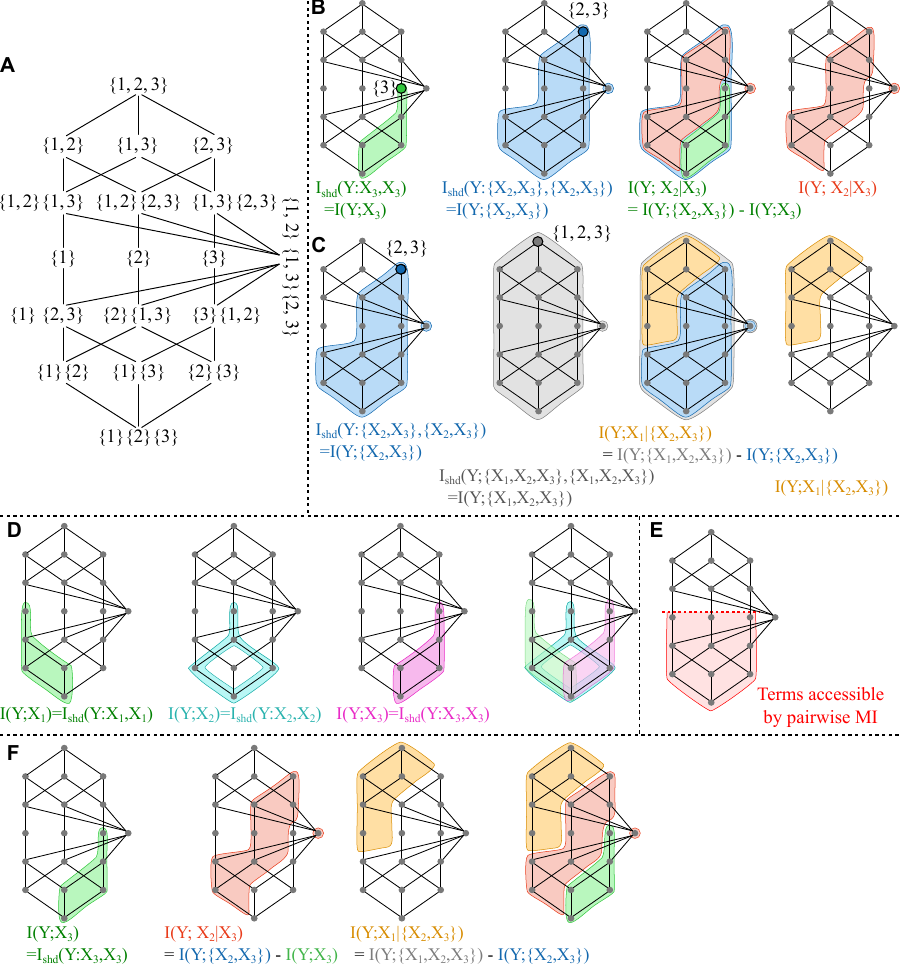}
    \caption{\textbf{Partial information decomposition (PID) redundancy lattice.}
    \textbf{A} PID lattice for four variables, i.e.,\ three features and one output variable. Sets indicate collections where numbers denote input variables $F_1$, $F_2$, $F_3$, while the output variable, $Y$, is implied via the fact that this information lattice represents a decomposition of $I(Y;F_1,F_2,F_3)$.
    \textbf{B} Information atoms covered by the conditional mutual information (CMI, red), $I(Y;F_2|F_3)=I(Y,\{F_2,F_3\})-I(Y,\{F_2\})$, derived from the complement of sets implied by the two mutual informations (MI), $I(Y;\{F_2,F_3\})$ (blue) and $I(Y;F_3)$ (green);
    \textbf{C} Information atoms covered by the CMI (yellow), $I(Y;F_1|\{F_2,F_3\})=I(Y,\{F_1,F_2,F_3\})-I(Y,\{F_2,F_3\})$, derived from the sets $I(Y,\{F_1,F_2,F_3\})$ (gray) and $I(Y,\{F_2,F_3\})$ (blue);
    \textbf{D} Information atoms covered when iteratively including using MI terms: many information atoms are covered multiple times meaning they are considered multiple times thus creating redundancies, higher-order information atoms are not covered.
    \textbf{E} Information atoms theoretically \enquote{accessible} when using MI (red area).
    \textbf{F} Information atoms covered when iteratively including using CMI terms: every collection is covered once, hence avoiding redundancies, also higher-order Information atoms are covered.
    }\label{fig:lattice}
\end{figure}

Figure~\ref{fig:lattice} shows the lattice for three features, $F_1$, $F_2$, $F_3$. The bottom node, $\{1\}\{2\}\{3\}$, indicates the information the three features share about the target $Y$. Nodes that contain more than one number within a bracket denote synergistic contributions, e.g., node $\{1,2\}$ denotes the synergistic contribution between features $F_1$ and $F_2$ with respect to $Y$, and the top node, $\{1,2,3\}$, represents information only accessible when knowing all feature variables together, i.e., it is the highest order synergy. A node containing a single number, e.g., $\{3\}$, denotes the information uniquely provided by feature $F_3$ about $Y$. Other nodes in the lattice, however, have a more complex interpretation. For a detailed introduction to parthood lattices and information atom interpretation see~\cite{Gutknecht2021}. Importantly, the MI between a single feature or a set of features and the target variable can be calculated from the node containing the feature in question and all child nodes (see, for example, the nodes that jointly represent $I(Y;F_3)$ and $I(Y;\{F2, F3\})$ in Figure~\ref{fig:lattice}B). By representing PID atoms in such a lattice structure, it can be shown that individual atoms can be calculated via a so-called Möbius Inversion from the redundant information terms and the classical mutual information terms.

In Figure~\ref{fig:lattice} we illustrate the information accessible by the MI versus the CMI for the case of four variables, i.e.,\ three features (for higher numbers of features our reasoning applies unchanged). Comparing the information atoms that are contained in the CMI terms computed in iterative selection (Figure~\ref{fig:lattice}B and C) to the information atoms contained in the corresponding MI terms (Figure~\ref{fig:lattice}D), one sees, that the atoms covered by the MI terms are highly overlapping (Figure~\ref{fig:lattice}D). In contrast, the atoms covered by the CMI terms cover the whole lattice non-redundantly (Figure~\ref{fig:lattice}F). Last, it is important to note that none of the unconditional MI terms covers atoms above the level of the individual unique information terms, $\{i\}$, on the lattice (Figure~\ref{fig:lattice}E)---in other words, none of the unconditional MI terms include higher-order interactions with respect to the target. The above observations all hold in general. For example, the inaccessibility of information above the level of the nodes carrying the unique information terms, $\{i\}$, by the unconditional MI, derives from the fact that these nodes are linked to the self-shared information, $I_{shd}(Y;F_i,F_i)=I(Y;F_i)$. Thus, the unconditional MI, $I(Y;F_i)$,---as a self-shared information---is limited to covering atoms below $\{i\}$ (also see~\citealt{Gutknecht2021} or~\citealt{Williams2010} for a more formal verification of this fact).

\section{Statistical Testing in Sequential Forward-Selection}\label{app:algo_stats}

To use the CMI or MI as a criterion in feature selection, estimation bias has to be handled such as to determine whether an estimate is truly non-zero. Estimators may return non-zero results also for independent variables due to finite sample size, and estimates may even be negative if the estimator bias is larger than the CMI \citep{Paninski2003,Kraskov2004,Hlavackova-Schindler2007}. The proposed sequential forward-selection algorithm handles estimation bias by performing non-parametric statistical tests against surrogate data to assess whether the estimate is statistically significant under the Null hypothesis of conditionally independent variables \citep{Novelli2019}. To generate the Null distribution, the CMI is repeatedly estimated from surrogate data, which is generated by permuting the data, in particular the target variable, $Y$, such that the joint distribution is destroyed while the marginal distributions stay intact.

During sequential forward-selection, statistical testing has to be performed repeatedly over all candidate features in each iteration. To control the family-wise error rate over these tests, a testing scheme termed \textit{maximum statistic} is used \citep{Novelli2019,Wollstadt2019}. The maximum statistic mimics the feature selection process by creating surrogate data from all remaining, non-selected variables in the current iteration, $\mathbf{X}_i = \mathbf{X}\setminus\mathbf{S}_i$. For each variable, $X\in\mathbf{X}_i$, realizations are permuted to generate surrogate data $X^\prime$, from which a surrogate value $I(X^\prime;Y|\mathbf{S}_i)$ is estimated. Then the maximum surrogate value over all variables $X\in\mathbf{X}_i$ is selected and stored. This  procedure is repeated a sufficient number of times to create  a distribution of maxima over surrogate estimates from all remaining, permuted variables. The test statistic (the CMI estimate from the original data) is then compared against the distribution of maxima (see~\citealt{Novelli2019} for a detailed description of the maximum statistic and a proof of its equivalence to the Dunn-\u{S}id\'ak correction, \citealt{Sidak1967}). For the final backward-elimination step, the same procedure is used, except that the minima over all surrogate values are collected because iteratively the feature with the \textit{minimal} contribution is tested.

By assessing whether the CMI is non-zero, statistical testing ensures the termination of the algorithm if none of the remaining variables adds information about the target in a significant fashion, thus providing an automatic stopping criterion and ensuring the selected feature set to be minimal \citep{Brown2012}.

\section{Feature Selection Problems of Medium Size}\label{app:medium_size_problem}

\begin{figure}[ht]
    \centering
    \includegraphics[width=1.0\textwidth]{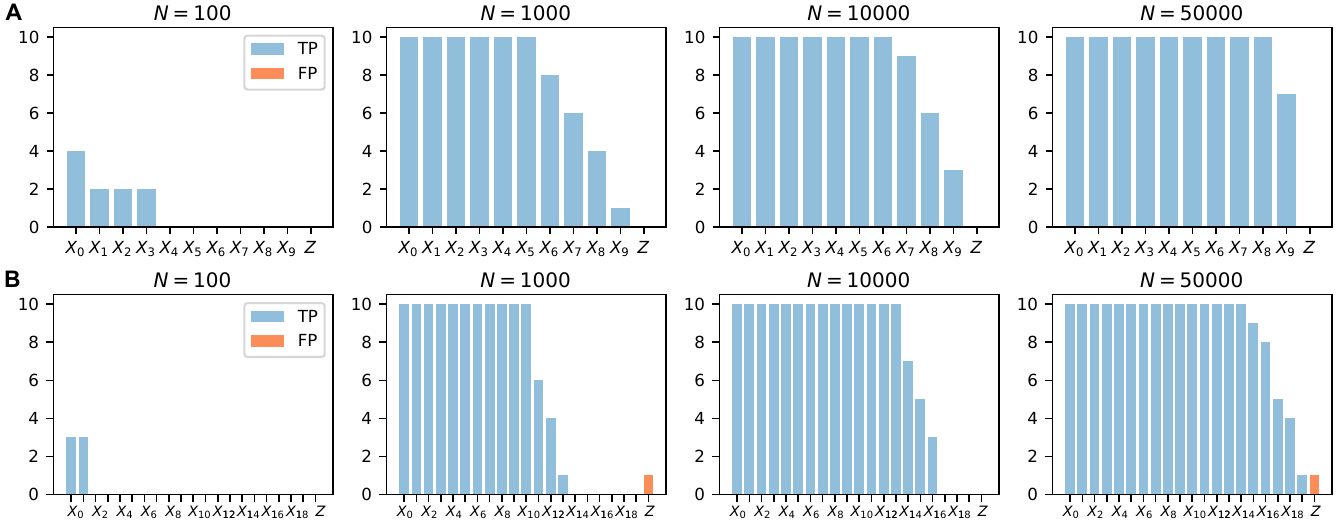}
    \caption{\textbf{Number of correctly identified features in larger-scale problems.}
    Total number of true positive results per informative feature (blue bars) and total number of false positives, $Z$ (orange bars), over 10 experimental runs. From left to right, results are shown for increasing sample size:
    \textbf{A} results for 10 informative features out of 50 variables;
    \textbf{B} results for 20 informative features out of 30 variables.
    }\label{fig:real_world_problem_size}
\end{figure}

To illustrate the algorithm's ability to handle feature selection problems of more realistic sizes, we generated a series of feature selection problems using scikit-learn \citep{scikit-learn} with varying numbers of variables, informative features, and sample sizes, $N$ (Fig.~\ref{fig:real_world_problem_size}). The target variable in each problem is generated via a linear regression model with the scikit-learn function \texttt{datasets.make\_regression}. We repeated each experiment 10 times. The algorithm was able to correctly identify half of all relevant features for a sample size of $N=1000$, and was able to retrieve 9 out of 10 relevant features for $N=50000$. For smaller samples size of $N=100$, only a small number of features was correctly identified in some of the runs. The number of false positives was well below the critical alpha level of $0.05$ used for the analysis.

In sum, the algorithm was able to reliably identify features if sample size was sufficient. For smaller sample sizes, the algorithm terminated before identifying all relevant features, which led to a higher number of false negatives, while the number of false positives was well below the level specified by the critical alpha level used in the statistical testing procedure. With an increasing number of samples, the algorithm was able to identify more relevant features correctly in both cases.

\vskip 0.2in
\bibliographystyle{apalike}
\bibliography{pid_feature_selection}

\end{document}